\documentclass[12pt,oneside,a4paper]{book} 
\usepackage{setspace} 

\usepackage[left=38mm,top=25mm,right=25mm,bottom=25mm]{geometry} 
\usepackage{amssymb}
\usepackage{color} 
\definecolor{AliceBlue}{rgb}{0.94,0.97,1.00}
\definecolor{AntiqueWhite1}{rgb}{1.00,0.94,0.86}
\definecolor{AntiqueWhite2}{rgb}{0.93,0.87,0.80}
\definecolor{AntiqueWhite3}{rgb}{0.80,0.75,0.69}
\definecolor{AntiqueWhite4}{rgb}{0.55,0.51,0.47}
\definecolor{AntiqueWhite}{rgb}{0.98,0.92,0.84}
\definecolor{BlanchedAlmond}{rgb}{1.00,0.92,0.80}
\definecolor{BlueViolet}{rgb}{0.54,0.17,0.89}
\definecolor{CadetBlue1}{rgb}{0.60,0.96,1.00}
\definecolor{CadetBlue2}{rgb}{0.56,0.90,0.93}
\definecolor{CadetBlue3}{rgb}{0.48,0.77,0.80}
\definecolor{CadetBlue4}{rgb}{0.33,0.53,0.55}
\definecolor{CadetBlue}{rgb}{0.37,0.62,0.63}
\definecolor{CornflowerBlue}{rgb}{0.39,0.58,0.93}
\definecolor{DarkBlue}{rgb}{0.00,0.00,0.55}
\definecolor{DarkCyan}{rgb}{0.00,0.55,0.55}
\definecolor{DarkGoldenrod1}{rgb}{1.00,0.73,0.06}
\definecolor{DarkGoldenrod2}{rgb}{0.93,0.68,0.05}
\definecolor{DarkGoldenrod3}{rgb}{0.80,0.58,0.05}
\definecolor{DarkGoldenrod4}{rgb}{0.55,0.40,0.03}
\definecolor{DarkGoldenrod}{rgb}{0.72,0.53,0.04}
\definecolor{DarkGray}{rgb}{0.66,0.66,0.66}
\definecolor{DarkGreen}{rgb}{0.00,0.39,0.00}
\definecolor{DarkGrey}{rgb}{0.66,0.66,0.66}
\definecolor{DarkKhaki}{rgb}{0.74,0.72,0.42}
\definecolor{DarkMagenta}{rgb}{0.55,0.00,0.55}
\definecolor{DarkOliveGreen1}{rgb}{0.79,1.00,0.44}
\definecolor{DarkOliveGreen2}{rgb}{0.74,0.93,0.41}
\definecolor{DarkOliveGreen3}{rgb}{0.64,0.80,0.35}
\definecolor{DarkOliveGreen4}{rgb}{0.43,0.55,0.24}
\definecolor{DarkOliveGreen}{rgb}{0.33,0.42,0.18}
\definecolor{DarkOrange1}{rgb}{1.00,0.50,0.00}
\definecolor{DarkOrange2}{rgb}{0.93,0.46,0.00}
\definecolor{DarkOrange3}{rgb}{0.80,0.40,0.00}
\definecolor{DarkOrange4}{rgb}{0.55,0.27,0.00}
\definecolor{DarkOrange}{rgb}{1.00,0.55,0.00}
\definecolor{DarkOrchid1}{rgb}{0.75,0.24,1.00}
\definecolor{DarkOrchid2}{rgb}{0.70,0.23,0.93}
\definecolor{DarkOrchid3}{rgb}{0.60,0.20,0.80}
\definecolor{DarkOrchid4}{rgb}{0.41,0.13,0.55}
\definecolor{DarkOrchid}{rgb}{0.60,0.20,0.80}
\definecolor{DarkRed}{rgb}{0.55,0.00,0.00}
\definecolor{DarkSalmon}{rgb}{0.91,0.59,0.48}
\definecolor{DarkSeaGreen1}{rgb}{0.76,1.00,0.76}
\definecolor{DarkSeaGreen2}{rgb}{0.71,0.93,0.71}
\definecolor{DarkSeaGreen3}{rgb}{0.61,0.80,0.61}
\definecolor{DarkSeaGreen4}{rgb}{0.41,0.55,0.41}
\definecolor{DarkSeaGreen}{rgb}{0.56,0.74,0.56}
\definecolor{DarkSlateBlue}{rgb}{0.28,0.24,0.55}
\definecolor{DarkSlateGray1}{rgb}{0.59,1.00,1.00}
\definecolor{DarkSlateGray2}{rgb}{0.55,0.93,0.93}
\definecolor{DarkSlateGray3}{rgb}{0.47,0.80,0.80}
\definecolor{DarkSlateGray4}{rgb}{0.32,0.55,0.55}
\definecolor{DarkSlateGray}{rgb}{0.18,0.31,0.31}
\definecolor{DarkSlateGrey}{rgb}{0.18,0.31,0.31}
\definecolor{DarkTurquoise}{rgb}{0.00,0.81,0.82}
\definecolor{DarkViolet}{rgb}{0.58,0.00,0.83}
\definecolor{DeepPink1}{rgb}{1.00,0.08,0.58}
\definecolor{DeepPink2}{rgb}{0.93,0.07,0.54}
\definecolor{DeepPink3}{rgb}{0.80,0.06,0.46}
\definecolor{DeepPink4}{rgb}{0.55,0.04,0.31}
\definecolor{DeepPink}{rgb}{1.00,0.08,0.58}
\definecolor{DeepSkyBlue1}{rgb}{0.00,0.75,1.00}
\definecolor{DeepSkyBlue2}{rgb}{0.00,0.70,0.93}
\definecolor{DeepSkyBlue3}{rgb}{0.00,0.60,0.80}
\definecolor{DeepSkyBlue4}{rgb}{0.00,0.41,0.55}
\definecolor{DeepSkyBlue}{rgb}{0.00,0.75,1.00}
\definecolor{DimGray}{rgb}{0.41,0.41,0.41}
\definecolor{DimGrey}{rgb}{0.41,0.41,0.41}
\definecolor{DodgerBlue1}{rgb}{0.12,0.56,1.00}
\definecolor{DodgerBlue2}{rgb}{0.11,0.53,0.93}
\definecolor{DodgerBlue3}{rgb}{0.09,0.45,0.80}
\definecolor{DodgerBlue4}{rgb}{0.06,0.31,0.55}
\definecolor{DodgerBlue}{rgb}{0.12,0.56,1.00}
\definecolor{FloralWhite}{rgb}{1.00,0.98,0.94}
\definecolor{ForestGreen}{rgb}{0.13,0.55,0.13}
\definecolor{GhostWhite}{rgb}{0.97,0.97,1.00}
\definecolor{GreenYellow}{rgb}{0.68,1.00,0.18}
\definecolor{HotPink1}{rgb}{1.00,0.43,0.71}
\definecolor{HotPink2}{rgb}{0.93,0.42,0.65}
\definecolor{HotPink3}{rgb}{0.80,0.38,0.56}
\definecolor{HotPink4}{rgb}{0.55,0.23,0.38}
\definecolor{HotPink}{rgb}{1.00,0.41,0.71}
\definecolor{IndianRed1}{rgb}{1.00,0.42,0.42}
\definecolor{IndianRed2}{rgb}{0.93,0.39,0.39}
\definecolor{IndianRed3}{rgb}{0.80,0.33,0.33}
\definecolor{IndianRed4}{rgb}{0.55,0.23,0.23}
\definecolor{IndianRed}{rgb}{0.80,0.36,0.36}
\definecolor{LavenderBlush1}{rgb}{1.00,0.94,0.96}
\definecolor{LavenderBlush2}{rgb}{0.93,0.88,0.90}
\definecolor{LavenderBlush3}{rgb}{0.80,0.76,0.77}
\definecolor{LavenderBlush4}{rgb}{0.55,0.51,0.53}
\definecolor{LavenderBlush}{rgb}{1.00,0.94,0.96}
\definecolor{LawnGreen}{rgb}{0.49,0.99,0.00}
\definecolor{LemonChiffon1}{rgb}{1.00,0.98,0.80}
\definecolor{LemonChiffon2}{rgb}{0.93,0.91,0.75}
\definecolor{LemonChiffon3}{rgb}{0.80,0.79,0.65}
\definecolor{LemonChiffon4}{rgb}{0.55,0.54,0.44}
\definecolor{LemonChiffon}{rgb}{1.00,0.98,0.80}
\definecolor{LightBlue1}{rgb}{0.75,0.94,1.00}
\definecolor{LightBlue2}{rgb}{0.70,0.87,0.93}
\definecolor{LightBlue3}{rgb}{0.60,0.75,0.80}
\definecolor{LightBlue4}{rgb}{0.41,0.51,0.55}
\definecolor{LightBlue}{rgb}{0.68,0.85,0.90}
\definecolor{LightCoral}{rgb}{0.94,0.50,0.50}
\definecolor{LightCyan1}{rgb}{0.88,1.00,1.00}
\definecolor{LightCyan2}{rgb}{0.82,0.93,0.93}
\definecolor{LightCyan3}{rgb}{0.71,0.80,0.80}
\definecolor{LightCyan4}{rgb}{0.48,0.55,0.55}
\definecolor{LightCyan}{rgb}{0.88,1.00,1.00}
\definecolor{LightGoldenrod1}{rgb}{1.00,0.93,0.55}
\definecolor{LightGoldenrod2}{rgb}{0.93,0.86,0.51}
\definecolor{LightGoldenrod3}{rgb}{0.80,0.75,0.44}
\definecolor{LightGoldenrod4}{rgb}{0.55,0.51,0.30}
\definecolor{LightGoldenrodYellow}{rgb}{0.98,0.98,0.82}
\definecolor{LightGoldenrod}{rgb}{0.93,0.87,0.51}
\definecolor{LightGray}{rgb}{0.83,0.83,0.83}
\definecolor{LightGreen}{rgb}{0.56,0.93,0.56}
\definecolor{LightGrey}{rgb}{0.83,0.83,0.83}
\definecolor{LightPink1}{rgb}{1.00,0.68,0.73}
\definecolor{LightPink2}{rgb}{0.93,0.64,0.68}
\definecolor{LightPink3}{rgb}{0.80,0.55,0.58}
\definecolor{LightPink4}{rgb}{0.55,0.37,0.40}
\definecolor{LightPink}{rgb}{1.00,0.71,0.76}
\definecolor{LightSalmon1}{rgb}{1.00,0.63,0.48}
\definecolor{LightSalmon2}{rgb}{0.93,0.58,0.45}
\definecolor{LightSalmon3}{rgb}{0.80,0.51,0.38}
\definecolor{LightSalmon4}{rgb}{0.55,0.34,0.26}
\definecolor{LightSalmon}{rgb}{1.00,0.63,0.48}
\definecolor{LightSeaGreen}{rgb}{0.13,0.70,0.67}
\definecolor{LightSkyBlue1}{rgb}{0.69,0.89,1.00}
\definecolor{LightSkyBlue2}{rgb}{0.64,0.83,0.93}
\definecolor{LightSkyBlue3}{rgb}{0.55,0.71,0.80}
\definecolor{LightSkyBlue4}{rgb}{0.38,0.48,0.55}
\definecolor{LightSkyBlue}{rgb}{0.53,0.81,0.98}
\definecolor{LightSlateBlue}{rgb}{0.52,0.44,1.00}
\definecolor{LightSlateGray}{rgb}{0.47,0.53,0.60}
\definecolor{LightSlateGrey}{rgb}{0.47,0.53,0.60}
\definecolor{LightSteelBlue1}{rgb}{0.79,0.88,1.00}
\definecolor{LightSteelBlue2}{rgb}{0.74,0.82,0.93}
\definecolor{LightSteelBlue3}{rgb}{0.64,0.71,0.80}
\definecolor{LightSteelBlue4}{rgb}{0.43,0.48,0.55}
\definecolor{LightSteelBlue}{rgb}{0.69,0.77,0.87}
\definecolor{LightYellow1}{rgb}{1.00,1.00,0.88}
\definecolor{LightYellow2}{rgb}{0.93,0.93,0.82}
\definecolor{LightYellow3}{rgb}{0.80,0.80,0.71}
\definecolor{LightYellow4}{rgb}{0.55,0.55,0.48}
\definecolor{LightYellow}{rgb}{1.00,1.00,0.88}
\definecolor{LimeGreen}{rgb}{0.20,0.80,0.20}
\definecolor{MediumAquamarine}{rgb}{0.40,0.80,0.67}
\definecolor{MediumBlue}{rgb}{0.00,0.00,0.80}
\definecolor{MediumOrchid1}{rgb}{0.88,0.40,1.00}
\definecolor{MediumOrchid2}{rgb}{0.82,0.37,0.93}
\definecolor{MediumOrchid3}{rgb}{0.71,0.32,0.80}
\definecolor{MediumOrchid4}{rgb}{0.48,0.22,0.55}
\definecolor{MediumOrchid}{rgb}{0.73,0.33,0.83}
\definecolor{MediumPurple1}{rgb}{0.67,0.51,1.00}
\definecolor{MediumPurple2}{rgb}{0.62,0.47,0.93}
\definecolor{MediumPurple3}{rgb}{0.54,0.41,0.80}
\definecolor{MediumPurple4}{rgb}{0.36,0.28,0.55}
\definecolor{MediumPurple}{rgb}{0.58,0.44,0.86}
\definecolor{MediumSeaGreen}{rgb}{0.24,0.70,0.44}
\definecolor{MediumSlateBlue}{rgb}{0.48,0.41,0.93}
\definecolor{MediumSpringGreen}{rgb}{0.00,0.98,0.60}
\definecolor{MediumTurquoise}{rgb}{0.28,0.82,0.80}
\definecolor{MediumVioletRed}{rgb}{0.78,0.08,0.52}
\definecolor{MidnightBlue}{rgb}{0.10,0.10,0.44}
\definecolor{MintCream}{rgb}{0.96,1.00,0.98}
\definecolor{MistyRose1}{rgb}{1.00,0.89,0.88}
\definecolor{MistyRose2}{rgb}{0.93,0.84,0.82}
\definecolor{MistyRose3}{rgb}{0.80,0.72,0.71}
\definecolor{MistyRose4}{rgb}{0.55,0.49,0.48}
\definecolor{MistyRose}{rgb}{1.00,0.89,0.88}
\definecolor{NavajoWhite1}{rgb}{1.00,0.87,0.68}
\definecolor{NavajoWhite2}{rgb}{0.93,0.81,0.63}
\definecolor{NavajoWhite3}{rgb}{0.80,0.70,0.55}
\definecolor{NavajoWhite4}{rgb}{0.55,0.47,0.37}
\definecolor{NavajoWhite}{rgb}{1.00,0.87,0.68}
\definecolor{NavyBlue}{rgb}{0.00,0.00,0.50}
\definecolor{OldLace}{rgb}{0.99,0.96,0.90}
\definecolor{OliveDrab1}{rgb}{0.75,1.00,0.24}
\definecolor{OliveDrab2}{rgb}{0.70,0.93,0.23}
\definecolor{OliveDrab3}{rgb}{0.60,0.80,0.20}
\definecolor{OliveDrab4}{rgb}{0.41,0.55,0.13}
\definecolor{OliveDrab}{rgb}{0.42,0.56,0.14}
\definecolor{OrangeRed1}{rgb}{1.00,0.27,0.00}
\definecolor{OrangeRed2}{rgb}{0.93,0.25,0.00}
\definecolor{OrangeRed3}{rgb}{0.80,0.22,0.00}
\definecolor{OrangeRed4}{rgb}{0.55,0.15,0.00}
\definecolor{OrangeRed}{rgb}{1.00,0.27,0.00}
\definecolor{PaleGoldenrod}{rgb}{0.93,0.91,0.67}
\definecolor{PaleGreen1}{rgb}{0.60,1.00,0.60}
\definecolor{PaleGreen2}{rgb}{0.56,0.93,0.56}
\definecolor{PaleGreen3}{rgb}{0.49,0.80,0.49}
\definecolor{PaleGreen4}{rgb}{0.33,0.55,0.33}
\definecolor{PaleGreen}{rgb}{0.60,0.98,0.60}
\definecolor{PaleTurquoise1}{rgb}{0.73,1.00,1.00}
\definecolor{PaleTurquoise2}{rgb}{0.68,0.93,0.93}
\definecolor{PaleTurquoise3}{rgb}{0.59,0.80,0.80}
\definecolor{PaleTurquoise4}{rgb}{0.40,0.55,0.55}
\definecolor{PaleTurquoise}{rgb}{0.69,0.93,0.93}
\definecolor{PaleVioletRed1}{rgb}{1.00,0.51,0.67}
\definecolor{PaleVioletRed2}{rgb}{0.93,0.47,0.62}
\definecolor{PaleVioletRed3}{rgb}{0.80,0.41,0.54}
\definecolor{PaleVioletRed4}{rgb}{0.55,0.28,0.36}
\definecolor{PaleVioletRed}{rgb}{0.86,0.44,0.58}
\definecolor{PapayaWhip}{rgb}{1.00,0.94,0.84}
\definecolor{PeachPuff1}{rgb}{1.00,0.85,0.73}
\definecolor{PeachPuff2}{rgb}{0.93,0.80,0.68}
\definecolor{PeachPuff3}{rgb}{0.80,0.69,0.58}
\definecolor{PeachPuff4}{rgb}{0.55,0.47,0.40}
\definecolor{PeachPuff}{rgb}{1.00,0.85,0.73}
\definecolor{PowderBlue}{rgb}{0.69,0.88,0.90}
\definecolor{RosyBrown1}{rgb}{1.00,0.76,0.76}
\definecolor{RosyBrown2}{rgb}{0.93,0.71,0.71}
\definecolor{RosyBrown3}{rgb}{0.80,0.61,0.61}
\definecolor{RosyBrown4}{rgb}{0.55,0.41,0.41}
\definecolor{RosyBrown}{rgb}{0.74,0.56,0.56}
\definecolor{RoyalBlue1}{rgb}{0.28,0.46,1.00}
\definecolor{RoyalBlue2}{rgb}{0.26,0.43,0.93}
\definecolor{RoyalBlue3}{rgb}{0.23,0.37,0.80}
\definecolor{RoyalBlue4}{rgb}{0.15,0.25,0.55}
\definecolor{RoyalBlue}{rgb}{0.25,0.41,0.88}
\definecolor{SaddleBrown}{rgb}{0.55,0.27,0.07}
\definecolor{SandyBrown}{rgb}{0.96,0.64,0.38}
\definecolor{SeaGreen1}{rgb}{0.33,1.00,0.62}
\definecolor{SeaGreen2}{rgb}{0.31,0.93,0.58}
\definecolor{SeaGreen3}{rgb}{0.26,0.80,0.50}
\definecolor{SeaGreen4}{rgb}{0.18,0.55,0.34}
\definecolor{SeaGreen}{rgb}{0.18,0.55,0.34}
\definecolor{SkyBlue1}{rgb}{0.53,0.81,1.00}
\definecolor{SkyBlue2}{rgb}{0.49,0.75,0.93}
\definecolor{SkyBlue3}{rgb}{0.42,0.65,0.80}
\definecolor{SkyBlue4}{rgb}{0.29,0.44,0.55}
\definecolor{SkyBlue}{rgb}{0.53,0.81,0.92}
\definecolor{SlateBlue1}{rgb}{0.51,0.44,1.00}
\definecolor{SlateBlue2}{rgb}{0.48,0.40,0.93}
\definecolor{SlateBlue3}{rgb}{0.41,0.35,0.80}
\definecolor{SlateBlue4}{rgb}{0.28,0.24,0.55}
\definecolor{SlateBlue}{rgb}{0.42,0.35,0.80}
\definecolor{SlateGray1}{rgb}{0.78,0.89,1.00}
\definecolor{SlateGray2}{rgb}{0.73,0.83,0.93}
\definecolor{SlateGray3}{rgb}{0.62,0.71,0.80}
\definecolor{SlateGray4}{rgb}{0.42,0.48,0.55}
\definecolor{SlateGray}{rgb}{0.44,0.50,0.56}
\definecolor{SlateGrey}{rgb}{0.44,0.50,0.56}
\definecolor{SpringGreen1}{rgb}{0.00,1.00,0.50}
\definecolor{SpringGreen2}{rgb}{0.00,0.93,0.46}
\definecolor{SpringGreen3}{rgb}{0.00,0.80,0.40}
\definecolor{SpringGreen4}{rgb}{0.00,0.55,0.27}
\definecolor{SpringGreen}{rgb}{0.00,1.00,0.50}
\definecolor{SteelBlue1}{rgb}{0.39,0.72,1.00}
\definecolor{SteelBlue2}{rgb}{0.36,0.67,0.93}
\definecolor{SteelBlue3}{rgb}{0.31,0.58,0.80}
\definecolor{SteelBlue4}{rgb}{0.21,0.39,0.55}
\definecolor{SteelBlue}{rgb}{0.27,0.51,0.71}
\definecolor{VioletRed1}{rgb}{1.00,0.24,0.59}
\definecolor{VioletRed2}{rgb}{0.93,0.23,0.55}
\definecolor{VioletRed3}{rgb}{0.80,0.20,0.47}
\definecolor{VioletRed4}{rgb}{0.55,0.13,0.32}
\definecolor{VioletRed}{rgb}{0.82,0.13,0.56}
\definecolor{WhiteSmoke}{rgb}{0.96,0.96,0.96}
\definecolor{YellowGreen}{rgb}{0.60,0.80,0.20}
\definecolor{aliceblue}{rgb}{0.94,0.97,1.00}
\definecolor{antiquewhite}{rgb}{0.98,0.92,0.84}
\definecolor{aquamarine1}{rgb}{0.50,1.00,0.83}
\definecolor{aquamarine2}{rgb}{0.46,0.93,0.78}
\definecolor{aquamarine3}{rgb}{0.40,0.80,0.67}
\definecolor{aquamarine4}{rgb}{0.27,0.55,0.45}
\definecolor{aquamarine}{rgb}{0.50,1.00,0.83}
\definecolor{azure1}{rgb}{0.94,1.00,1.00}
\definecolor{azure2}{rgb}{0.88,0.93,0.93}
\definecolor{azure3}{rgb}{0.76,0.80,0.80}
\definecolor{azure4}{rgb}{0.51,0.55,0.55}
\definecolor{azure}{rgb}{0.94,1.00,1.00}
\definecolor{beige}{rgb}{0.96,0.96,0.86}
\definecolor{bisque1}{rgb}{1.00,0.89,0.77}
\definecolor{bisque2}{rgb}{0.93,0.84,0.72}
\definecolor{bisque3}{rgb}{0.80,0.72,0.62}
\definecolor{bisque4}{rgb}{0.55,0.49,0.42}
\definecolor{bisque}{rgb}{1.00,0.89,0.77}
\definecolor{black}{rgb}{0.00,0.00,0.00}
\definecolor{blanchedalmond}{rgb}{1.00,0.92,0.80}
\definecolor{blue1}{rgb}{0.00,0.00,1.00}
\definecolor{blue2}{rgb}{0.00,0.00,0.93}
\definecolor{blue3}{rgb}{0.00,0.00,0.80}
\definecolor{blue4}{rgb}{0.00,0.00,0.55}
\definecolor{blueviolet}{rgb}{0.54,0.17,0.89}
\definecolor{blue}{rgb}{0.00,0.00,1.00}
\definecolor{brown1}{rgb}{1.00,0.25,0.25}
\definecolor{brown2}{rgb}{0.93,0.23,0.23}
\definecolor{brown3}{rgb}{0.80,0.20,0.20}
\definecolor{brown4}{rgb}{0.55,0.14,0.14}
\definecolor{brown}{rgb}{0.65,0.16,0.16}
\definecolor{burlywood1}{rgb}{1.00,0.83,0.61}
\definecolor{burlywood2}{rgb}{0.93,0.77,0.57}
\definecolor{burlywood3}{rgb}{0.80,0.67,0.49}
\definecolor{burlywood4}{rgb}{0.55,0.45,0.33}
\definecolor{burlywood}{rgb}{0.87,0.72,0.53}
\definecolor{cadetblue}{rgb}{0.37,0.62,0.63}
\definecolor{chartreuse1}{rgb}{0.50,1.00,0.00}
\definecolor{chartreuse2}{rgb}{0.46,0.93,0.00}
\definecolor{chartreuse3}{rgb}{0.40,0.80,0.00}
\definecolor{chartreuse4}{rgb}{0.27,0.55,0.00}
\definecolor{chartreuse}{rgb}{0.50,1.00,0.00}
\definecolor{chocolate1}{rgb}{1.00,0.50,0.14}
\definecolor{chocolate2}{rgb}{0.93,0.46,0.13}
\definecolor{chocolate3}{rgb}{0.80,0.40,0.11}
\definecolor{chocolate4}{rgb}{0.55,0.27,0.07}
\definecolor{chocolate}{rgb}{0.82,0.41,0.12}
\definecolor{coral1}{rgb}{1.00,0.45,0.34}
\definecolor{coral2}{rgb}{0.93,0.42,0.31}
\definecolor{coral3}{rgb}{0.80,0.36,0.27}
\definecolor{coral4}{rgb}{0.55,0.24,0.18}
\definecolor{coral}{rgb}{1.00,0.50,0.31}
\definecolor{cornflowerblue}{rgb}{0.39,0.58,0.93}
\definecolor{cornsilk1}{rgb}{1.00,0.97,0.86}
\definecolor{cornsilk2}{rgb}{0.93,0.91,0.80}
\definecolor{cornsilk3}{rgb}{0.80,0.78,0.69}
\definecolor{cornsilk4}{rgb}{0.55,0.53,0.47}
\definecolor{cornsilk}{rgb}{1.00,0.97,0.86}
\definecolor{cyan1}{rgb}{0.00,1.00,1.00}
\definecolor{cyan2}{rgb}{0.00,0.93,0.93}
\definecolor{cyan3}{rgb}{0.00,0.80,0.80}
\definecolor{cyan4}{rgb}{0.00,0.55,0.55}
\definecolor{cyan}{rgb}{0.00,1.00,1.00}
\definecolor{darkblue}{rgb}{0.00,0.00,0.55}
\definecolor{darkcyan}{rgb}{0.00,0.55,0.55}
\definecolor{darkgoldenrod}{rgb}{0.72,0.53,0.04}
\definecolor{darkgray}{rgb}{0.66,0.66,0.66}
\definecolor{darkgreen}{rgb}{0.00,0.39,0.00}
\definecolor{darkgrey}{rgb}{0.66,0.66,0.66}
\definecolor{darkkhaki}{rgb}{0.74,0.72,0.42}
\definecolor{darkmagenta}{rgb}{0.55,0.00,0.55}
\definecolor{darkolive}{rgb}{0.33,0.42,0.18}
\definecolor{darkorange}{rgb}{1.00,0.55,0.00}
\definecolor{darkorchid}{rgb}{0.60,0.20,0.80}
\definecolor{darkred}{rgb}{0.55,0.00,0.00}
\definecolor{darksalmon}{rgb}{0.91,0.59,0.48}
\definecolor{darksea}{rgb}{0.56,0.74,0.56}
\definecolor{darkslate}{rgb}{0.18,0.31,0.31}
\definecolor{darkslate}{rgb}{0.18,0.31,0.31}
\definecolor{darkslate}{rgb}{0.28,0.24,0.55}
\definecolor{darkturquoise}{rgb}{0.00,0.81,0.82}
\definecolor{darkviolet}{rgb}{0.58,0.00,0.83}
\definecolor{deeppink}{rgb}{1.00,0.08,0.58}
\definecolor{deepsky}{rgb}{0.00,0.75,1.00}
\definecolor{dimgray}{rgb}{0.41,0.41,0.41}
\definecolor{dimgrey}{rgb}{0.41,0.41,0.41}
\definecolor{dodgerblue}{rgb}{0.12,0.56,1.00}
\definecolor{firebrick1}{rgb}{1.00,0.19,0.19}
\definecolor{firebrick2}{rgb}{0.93,0.17,0.17}
\definecolor{firebrick3}{rgb}{0.80,0.15,0.15}
\definecolor{firebrick4}{rgb}{0.55,0.10,0.10}
\definecolor{firebrick}{rgb}{0.70,0.13,0.13}
\definecolor{floralwhite}{rgb}{1.00,0.98,0.94}
\definecolor{forestgreen}{rgb}{0.13,0.55,0.13}
\definecolor{gainsboro}{rgb}{0.86,0.86,0.86}
\definecolor{ghostwhite}{rgb}{0.97,0.97,1.00}
\definecolor{gold1}{rgb}{1.00,0.84,0.00}
\definecolor{gold2}{rgb}{0.93,0.79,0.00}
\definecolor{gold3}{rgb}{0.80,0.68,0.00}
\definecolor{gold4}{rgb}{0.55,0.46,0.00}
\definecolor{goldenrod1}{rgb}{1.00,0.76,0.15}
\definecolor{goldenrod2}{rgb}{0.93,0.71,0.13}
\definecolor{goldenrod3}{rgb}{0.80,0.61,0.11}
\definecolor{goldenrod4}{rgb}{0.55,0.41,0.08}
\definecolor{goldenrod}{rgb}{0.85,0.65,0.13}
\definecolor{gold}{rgb}{1.00,0.84,0.00}
\definecolor{gray0}{rgb}{0.00,0.00,0.00}
\definecolor{gray100}{rgb}{1.00,1.00,1.00}
\definecolor{gray10}{rgb}{0.10,0.10,0.10}
\definecolor{gray11}{rgb}{0.11,0.11,0.11}
\definecolor{gray12}{rgb}{0.12,0.12,0.12}
\definecolor{gray13}{rgb}{0.13,0.13,0.13}
\definecolor{gray14}{rgb}{0.14,0.14,0.14}
\definecolor{gray15}{rgb}{0.15,0.15,0.15}
\definecolor{gray16}{rgb}{0.16,0.16,0.16}
\definecolor{gray17}{rgb}{0.17,0.17,0.17}
\definecolor{gray18}{rgb}{0.18,0.18,0.18}
\definecolor{gray19}{rgb}{0.19,0.19,0.19}
\definecolor{gray1}{rgb}{0.01,0.01,0.01}
\definecolor{gray20}{rgb}{0.20,0.20,0.20}
\definecolor{gray21}{rgb}{0.21,0.21,0.21}
\definecolor{gray22}{rgb}{0.22,0.22,0.22}
\definecolor{gray23}{rgb}{0.23,0.23,0.23}
\definecolor{gray24}{rgb}{0.24,0.24,0.24}
\definecolor{gray25}{rgb}{0.25,0.25,0.25}
\definecolor{gray26}{rgb}{0.26,0.26,0.26}
\definecolor{gray27}{rgb}{0.27,0.27,0.27}
\definecolor{gray28}{rgb}{0.28,0.28,0.28}
\definecolor{gray29}{rgb}{0.29,0.29,0.29}
\definecolor{gray2}{rgb}{0.02,0.02,0.02}
\definecolor{gray30}{rgb}{0.30,0.30,0.30}
\definecolor{gray31}{rgb}{0.31,0.31,0.31}
\definecolor{gray32}{rgb}{0.32,0.32,0.32}
\definecolor{gray33}{rgb}{0.33,0.33,0.33}
\definecolor{gray34}{rgb}{0.34,0.34,0.34}
\definecolor{gray35}{rgb}{0.35,0.35,0.35}
\definecolor{gray36}{rgb}{0.36,0.36,0.36}
\definecolor{gray37}{rgb}{0.37,0.37,0.37}
\definecolor{gray38}{rgb}{0.38,0.38,0.38}
\definecolor{gray39}{rgb}{0.39,0.39,0.39}
\definecolor{gray3}{rgb}{0.03,0.03,0.03}
\definecolor{gray40}{rgb}{0.40,0.40,0.40}
\definecolor{gray41}{rgb}{0.41,0.41,0.41}
\definecolor{gray42}{rgb}{0.42,0.42,0.42}
\definecolor{gray43}{rgb}{0.43,0.43,0.43}
\definecolor{gray44}{rgb}{0.44,0.44,0.44}
\definecolor{gray45}{rgb}{0.45,0.45,0.45}
\definecolor{gray46}{rgb}{0.46,0.46,0.46}
\definecolor{gray47}{rgb}{0.47,0.47,0.47}
\definecolor{gray48}{rgb}{0.48,0.48,0.48}
\definecolor{gray49}{rgb}{0.49,0.49,0.49}
\definecolor{gray4}{rgb}{0.04,0.04,0.04}
\definecolor{gray50}{rgb}{0.50,0.50,0.50}
\definecolor{gray51}{rgb}{0.51,0.51,0.51}
\definecolor{gray52}{rgb}{0.52,0.52,0.52}
\definecolor{gray53}{rgb}{0.53,0.53,0.53}
\definecolor{gray54}{rgb}{0.54,0.54,0.54}
\definecolor{gray55}{rgb}{0.55,0.55,0.55}
\definecolor{gray56}{rgb}{0.56,0.56,0.56}
\definecolor{gray57}{rgb}{0.57,0.57,0.57}
\definecolor{gray58}{rgb}{0.58,0.58,0.58}
\definecolor{gray59}{rgb}{0.59,0.59,0.59}
\definecolor{gray5}{rgb}{0.05,0.05,0.05}
\definecolor{gray60}{rgb}{0.60,0.60,0.60}
\definecolor{gray61}{rgb}{0.61,0.61,0.61}
\definecolor{gray62}{rgb}{0.62,0.62,0.62}
\definecolor{gray63}{rgb}{0.63,0.63,0.63}
\definecolor{gray64}{rgb}{0.64,0.64,0.64}
\definecolor{gray65}{rgb}{0.65,0.65,0.65}
\definecolor{gray66}{rgb}{0.66,0.66,0.66}
\definecolor{gray67}{rgb}{0.67,0.67,0.67}
\definecolor{gray68}{rgb}{0.68,0.68,0.68}
\definecolor{gray69}{rgb}{0.69,0.69,0.69}
\definecolor{gray6}{rgb}{0.06,0.06,0.06}
\definecolor{gray70}{rgb}{0.70,0.70,0.70}
\definecolor{gray71}{rgb}{0.71,0.71,0.71}
\definecolor{gray72}{rgb}{0.72,0.72,0.72}
\definecolor{gray73}{rgb}{0.73,0.73,0.73}
\definecolor{gray74}{rgb}{0.74,0.74,0.74}
\definecolor{gray75}{rgb}{0.75,0.75,0.75}
\definecolor{gray76}{rgb}{0.76,0.76,0.76}
\definecolor{gray77}{rgb}{0.77,0.77,0.77}
\definecolor{gray78}{rgb}{0.78,0.78,0.78}
\definecolor{gray79}{rgb}{0.79,0.79,0.79}
\definecolor{gray7}{rgb}{0.07,0.07,0.07}
\definecolor{gray80}{rgb}{0.80,0.80,0.80}
\definecolor{gray81}{rgb}{0.81,0.81,0.81}
\definecolor{gray82}{rgb}{0.82,0.82,0.82}
\definecolor{gray83}{rgb}{0.83,0.83,0.83}
\definecolor{gray84}{rgb}{0.84,0.84,0.84}
\definecolor{gray85}{rgb}{0.85,0.85,0.85}
\definecolor{gray86}{rgb}{0.86,0.86,0.86}
\definecolor{gray87}{rgb}{0.87,0.87,0.87}
\definecolor{gray88}{rgb}{0.88,0.88,0.88}
\definecolor{gray89}{rgb}{0.89,0.89,0.89}
\definecolor{gray8}{rgb}{0.08,0.08,0.08}
\definecolor{gray90}{rgb}{0.90,0.90,0.90}
\definecolor{gray91}{rgb}{0.91,0.91,0.91}
\definecolor{gray92}{rgb}{0.92,0.92,0.92}
\definecolor{gray93}{rgb}{0.93,0.93,0.93}
\definecolor{gray94}{rgb}{0.94,0.94,0.94}
\definecolor{gray95}{rgb}{0.95,0.95,0.95}
\definecolor{gray96}{rgb}{0.96,0.96,0.96}
\definecolor{gray97}{rgb}{0.97,0.97,0.97}
\definecolor{gray98}{rgb}{0.98,0.98,0.98}
\definecolor{gray99}{rgb}{0.99,0.99,0.99}
\definecolor{gray9}{rgb}{0.09,0.09,0.09}
\definecolor{gray}{rgb}{0.75,0.75,0.75}
\definecolor{green1}{rgb}{0.00,1.00,0.00}
\definecolor{green2}{rgb}{0.00,0.93,0.00}
\definecolor{green3}{rgb}{0.00,0.80,0.00}
\definecolor{green4}{rgb}{0.00,0.55,0.00}
\definecolor{greenyellow}{rgb}{0.68,1.00,0.18}
\definecolor{green}{rgb}{0.00,1.00,0.00}
\definecolor{grey0}{rgb}{0.00,0.00,0.00}
\definecolor{grey100}{rgb}{1.00,1.00,1.00}
\definecolor{grey10}{rgb}{0.10,0.10,0.10}
\definecolor{grey11}{rgb}{0.11,0.11,0.11}
\definecolor{grey12}{rgb}{0.12,0.12,0.12}
\definecolor{grey13}{rgb}{0.13,0.13,0.13}
\definecolor{grey14}{rgb}{0.14,0.14,0.14}
\definecolor{grey15}{rgb}{0.15,0.15,0.15}
\definecolor{grey16}{rgb}{0.16,0.16,0.16}
\definecolor{grey17}{rgb}{0.17,0.17,0.17}
\definecolor{grey18}{rgb}{0.18,0.18,0.18}
\definecolor{grey19}{rgb}{0.19,0.19,0.19}
\definecolor{grey1}{rgb}{0.01,0.01,0.01}
\definecolor{grey20}{rgb}{0.20,0.20,0.20}
\definecolor{grey21}{rgb}{0.21,0.21,0.21}
\definecolor{grey22}{rgb}{0.22,0.22,0.22}
\definecolor{grey23}{rgb}{0.23,0.23,0.23}
\definecolor{grey24}{rgb}{0.24,0.24,0.24}
\definecolor{grey25}{rgb}{0.25,0.25,0.25}
\definecolor{grey26}{rgb}{0.26,0.26,0.26}
\definecolor{grey27}{rgb}{0.27,0.27,0.27}
\definecolor{grey28}{rgb}{0.28,0.28,0.28}
\definecolor{grey29}{rgb}{0.29,0.29,0.29}
\definecolor{grey2}{rgb}{0.02,0.02,0.02}
\definecolor{grey30}{rgb}{0.30,0.30,0.30}
\definecolor{grey31}{rgb}{0.31,0.31,0.31}
\definecolor{grey32}{rgb}{0.32,0.32,0.32}
\definecolor{grey33}{rgb}{0.33,0.33,0.33}
\definecolor{grey34}{rgb}{0.34,0.34,0.34}
\definecolor{grey35}{rgb}{0.35,0.35,0.35}
\definecolor{grey36}{rgb}{0.36,0.36,0.36}
\definecolor{grey37}{rgb}{0.37,0.37,0.37}
\definecolor{grey38}{rgb}{0.38,0.38,0.38}
\definecolor{grey39}{rgb}{0.39,0.39,0.39}
\definecolor{grey3}{rgb}{0.03,0.03,0.03}
\definecolor{grey40}{rgb}{0.40,0.40,0.40}
\definecolor{grey41}{rgb}{0.41,0.41,0.41}
\definecolor{grey42}{rgb}{0.42,0.42,0.42}
\definecolor{grey43}{rgb}{0.43,0.43,0.43}
\definecolor{grey44}{rgb}{0.44,0.44,0.44}
\definecolor{grey45}{rgb}{0.45,0.45,0.45}
\definecolor{grey46}{rgb}{0.46,0.46,0.46}
\definecolor{grey47}{rgb}{0.47,0.47,0.47}
\definecolor{grey48}{rgb}{0.48,0.48,0.48}
\definecolor{grey49}{rgb}{0.49,0.49,0.49}
\definecolor{grey4}{rgb}{0.04,0.04,0.04}
\definecolor{grey50}{rgb}{0.50,0.50,0.50}
\definecolor{grey51}{rgb}{0.51,0.51,0.51}
\definecolor{grey52}{rgb}{0.52,0.52,0.52}
\definecolor{grey53}{rgb}{0.53,0.53,0.53}
\definecolor{grey54}{rgb}{0.54,0.54,0.54}
\definecolor{grey55}{rgb}{0.55,0.55,0.55}
\definecolor{grey56}{rgb}{0.56,0.56,0.56}
\definecolor{grey57}{rgb}{0.57,0.57,0.57}
\definecolor{grey58}{rgb}{0.58,0.58,0.58}
\definecolor{grey59}{rgb}{0.59,0.59,0.59}
\definecolor{grey5}{rgb}{0.05,0.05,0.05}
\definecolor{grey60}{rgb}{0.60,0.60,0.60}
\definecolor{grey61}{rgb}{0.61,0.61,0.61}
\definecolor{grey62}{rgb}{0.62,0.62,0.62}
\definecolor{grey63}{rgb}{0.63,0.63,0.63}
\definecolor{grey64}{rgb}{0.64,0.64,0.64}
\definecolor{grey65}{rgb}{0.65,0.65,0.65}
\definecolor{grey66}{rgb}{0.66,0.66,0.66}
\definecolor{grey67}{rgb}{0.67,0.67,0.67}
\definecolor{grey68}{rgb}{0.68,0.68,0.68}
\definecolor{grey69}{rgb}{0.69,0.69,0.69}
\definecolor{grey6}{rgb}{0.06,0.06,0.06}
\definecolor{grey70}{rgb}{0.70,0.70,0.70}
\definecolor{grey71}{rgb}{0.71,0.71,0.71}
\definecolor{grey72}{rgb}{0.72,0.72,0.72}
\definecolor{grey73}{rgb}{0.73,0.73,0.73}
\definecolor{grey74}{rgb}{0.74,0.74,0.74}
\definecolor{grey75}{rgb}{0.75,0.75,0.75}
\definecolor{grey76}{rgb}{0.76,0.76,0.76}
\definecolor{grey77}{rgb}{0.77,0.77,0.77}
\definecolor{grey78}{rgb}{0.78,0.78,0.78}
\definecolor{grey79}{rgb}{0.79,0.79,0.79}
\definecolor{grey7}{rgb}{0.07,0.07,0.07}
\definecolor{grey80}{rgb}{0.80,0.80,0.80}
\definecolor{grey81}{rgb}{0.81,0.81,0.81}
\definecolor{grey82}{rgb}{0.82,0.82,0.82}
\definecolor{grey83}{rgb}{0.83,0.83,0.83}
\definecolor{grey84}{rgb}{0.84,0.84,0.84}
\definecolor{grey85}{rgb}{0.85,0.85,0.85}
\definecolor{grey86}{rgb}{0.86,0.86,0.86}
\definecolor{grey87}{rgb}{0.87,0.87,0.87}
\definecolor{grey88}{rgb}{0.88,0.88,0.88}
\definecolor{grey89}{rgb}{0.89,0.89,0.89}
\definecolor{grey8}{rgb}{0.08,0.08,0.08}
\definecolor{grey90}{rgb}{0.90,0.90,0.90}
\definecolor{grey91}{rgb}{0.91,0.91,0.91}
\definecolor{grey92}{rgb}{0.92,0.92,0.92}
\definecolor{grey93}{rgb}{0.93,0.93,0.93}
\definecolor{grey94}{rgb}{0.94,0.94,0.94}
\definecolor{grey95}{rgb}{0.95,0.95,0.95}
\definecolor{grey96}{rgb}{0.96,0.96,0.96}
\definecolor{grey97}{rgb}{0.97,0.97,0.97}
\definecolor{grey98}{rgb}{0.98,0.98,0.98}
\definecolor{grey99}{rgb}{0.99,0.99,0.99}
\definecolor{grey9}{rgb}{0.09,0.09,0.09}
\definecolor{grey}{rgb}{0.75,0.75,0.75}
\definecolor{honeydew1}{rgb}{0.94,1.00,0.94}
\definecolor{honeydew2}{rgb}{0.88,0.93,0.88}
\definecolor{honeydew3}{rgb}{0.76,0.80,0.76}
\definecolor{honeydew4}{rgb}{0.51,0.55,0.51}
\definecolor{honeydew}{rgb}{0.94,1.00,0.94}
\definecolor{hotpink}{rgb}{1.00,0.41,0.71}
\definecolor{indianred}{rgb}{0.80,0.36,0.36}
\definecolor{ivory1}{rgb}{1.00,1.00,0.94}
\definecolor{ivory2}{rgb}{0.93,0.93,0.88}
\definecolor{ivory3}{rgb}{0.80,0.80,0.76}
\definecolor{ivory4}{rgb}{0.55,0.55,0.51}
\definecolor{ivory}{rgb}{1.00,1.00,0.94}
\definecolor{khaki1}{rgb}{1.00,0.96,0.56}
\definecolor{khaki2}{rgb}{0.93,0.90,0.52}
\definecolor{khaki3}{rgb}{0.80,0.78,0.45}
\definecolor{khaki4}{rgb}{0.55,0.53,0.31}
\definecolor{khaki}{rgb}{0.94,0.90,0.55}
\definecolor{lavenderblush}{rgb}{1.00,0.94,0.96}
\definecolor{lavender}{rgb}{0.90,0.90,0.98}
\definecolor{lawngreen}{rgb}{0.49,0.99,0.00}
\definecolor{lemonchiffon}{rgb}{1.00,0.98,0.80}
\definecolor{lightblue}{rgb}{0.68,0.85,0.90}
\definecolor{lightcoral}{rgb}{0.94,0.50,0.50}
\definecolor{lightcyan}{rgb}{0.88,1.00,1.00}
\definecolor{lightgoldenrod}{rgb}{0.93,0.87,0.51}
\definecolor{lightgoldenrod}{rgb}{0.98,0.98,0.82}
\definecolor{lightgray}{rgb}{0.83,0.83,0.83}
\definecolor{lightgreen}{rgb}{0.56,0.93,0.56}
\definecolor{lightgrey}{rgb}{0.83,0.83,0.83}
\definecolor{lightpink}{rgb}{1.00,0.71,0.76}
\definecolor{lightsalmon}{rgb}{1.00,0.63,0.48}
\definecolor{lightsea}{rgb}{0.13,0.70,0.67}
\definecolor{lightsky}{rgb}{0.53,0.81,0.98}
\definecolor{lightslate}{rgb}{0.47,0.53,0.60}
\definecolor{lightslate}{rgb}{0.47,0.53,0.60}
\definecolor{lightslate}{rgb}{0.52,0.44,1.00}
\definecolor{lightsteel}{rgb}{0.69,0.77,0.87}
\definecolor{lightyellow}{rgb}{1.00,1.00,0.88}
\definecolor{limegreen}{rgb}{0.20,0.80,0.20}
\definecolor{linen}{rgb}{0.98,0.94,0.90}
\definecolor{magenta1}{rgb}{1.00,0.00,1.00}
\definecolor{magenta2}{rgb}{0.93,0.00,0.93}
\definecolor{magenta3}{rgb}{0.80,0.00,0.80}
\definecolor{magenta4}{rgb}{0.55,0.00,0.55}
\definecolor{magenta}{rgb}{1.00,0.00,1.00}
\definecolor{maroon1}{rgb}{1.00,0.20,0.70}
\definecolor{maroon2}{rgb}{0.93,0.19,0.65}
\definecolor{maroon3}{rgb}{0.80,0.16,0.56}
\definecolor{maroon4}{rgb}{0.55,0.11,0.38}
\definecolor{maroon}{rgb}{0.69,0.19,0.38}
\definecolor{mediumaquamarine}{rgb}{0.40,0.80,0.67}
\definecolor{mediumblue}{rgb}{0.00,0.00,0.80}
\definecolor{mediumorchid}{rgb}{0.73,0.33,0.83}
\definecolor{mediumpurple}{rgb}{0.58,0.44,0.86}
\definecolor{mediumsea}{rgb}{0.24,0.70,0.44}
\definecolor{mediumslate}{rgb}{0.48,0.41,0.93}
\definecolor{mediumspring}{rgb}{0.00,0.98,0.60}
\definecolor{mediumturquoise}{rgb}{0.28,0.82,0.80}
\definecolor{mediumviolet}{rgb}{0.78,0.08,0.52}
\definecolor{midnightblue}{rgb}{0.10,0.10,0.44}
\definecolor{mintcream}{rgb}{0.96,1.00,0.98}
\definecolor{mistyrose}{rgb}{1.00,0.89,0.88}
\definecolor{moccasin}{rgb}{1.00,0.89,0.71}
\definecolor{navajowhite}{rgb}{1.00,0.87,0.68}
\definecolor{navyblue}{rgb}{0.00,0.00,0.50}
\definecolor{navy}{rgb}{0.00,0.00,0.50}
\definecolor{oldlace}{rgb}{0.99,0.96,0.90}
\definecolor{olivedrab}{rgb}{0.42,0.56,0.14}
\definecolor{orange1}{rgb}{1.00,0.65,0.00}
\definecolor{orange2}{rgb}{0.93,0.60,0.00}
\definecolor{orange3}{rgb}{0.80,0.52,0.00}
\definecolor{orange4}{rgb}{0.55,0.35,0.00}
\definecolor{orangered}{rgb}{1.00,0.27,0.00}
\definecolor{orange}{rgb}{1.00,0.65,0.00}
\definecolor{orchid1}{rgb}{1.00,0.51,0.98}
\definecolor{orchid2}{rgb}{0.93,0.48,0.91}
\definecolor{orchid3}{rgb}{0.80,0.41,0.79}
\definecolor{orchid4}{rgb}{0.55,0.28,0.54}
\definecolor{orchid}{rgb}{0.85,0.44,0.84}
\definecolor{palegoldenrod}{rgb}{0.93,0.91,0.67}
\definecolor{palegreen}{rgb}{0.60,0.98,0.60}
\definecolor{paleturquoise}{rgb}{0.69,0.93,0.93}
\definecolor{paleviolet}{rgb}{0.86,0.44,0.58}
\definecolor{papayawhip}{rgb}{1.00,0.94,0.84}
\definecolor{peachpuff}{rgb}{1.00,0.85,0.73}
\definecolor{peru}{rgb}{0.80,0.52,0.25}
\definecolor{pink1}{rgb}{1.00,0.71,0.77}
\definecolor{pink2}{rgb}{0.93,0.66,0.72}
\definecolor{pink3}{rgb}{0.80,0.57,0.62}
\definecolor{pink4}{rgb}{0.55,0.39,0.42}
\definecolor{pink}{rgb}{1.00,0.75,0.80}
\definecolor{plum1}{rgb}{1.00,0.73,1.00}
\definecolor{plum2}{rgb}{0.93,0.68,0.93}
\definecolor{plum3}{rgb}{0.80,0.59,0.80}
\definecolor{plum4}{rgb}{0.55,0.40,0.55}
\definecolor{plum}{rgb}{0.87,0.63,0.87}
\definecolor{powderblue}{rgb}{0.69,0.88,0.90}
\definecolor{purple1}{rgb}{0.61,0.19,1.00}
\definecolor{purple2}{rgb}{0.57,0.17,0.93}
\definecolor{purple3}{rgb}{0.49,0.15,0.80}
\definecolor{purple4}{rgb}{0.33,0.10,0.55}
\definecolor{purple}{rgb}{0.63,0.13,0.94}
\definecolor{red1}{rgb}{1.00,0.00,0.00}
\definecolor{red2}{rgb}{0.93,0.00,0.00}
\definecolor{red3}{rgb}{0.80,0.00,0.00}
\definecolor{red4}{rgb}{0.55,0.00,0.00}
\definecolor{red}{rgb}{1.00,0.00,0.00}
\definecolor{rosybrown}{rgb}{0.74,0.56,0.56}
\definecolor{royalblue}{rgb}{0.25,0.41,0.88}
\definecolor{saddlebrown}{rgb}{0.55,0.27,0.07}
\definecolor{salmon1}{rgb}{1.00,0.55,0.41}
\definecolor{salmon2}{rgb}{0.93,0.51,0.38}
\definecolor{salmon3}{rgb}{0.80,0.44,0.33}
\definecolor{salmon4}{rgb}{0.55,0.30,0.22}
\definecolor{salmon}{rgb}{0.98,0.50,0.45}
\definecolor{sandybrown}{rgb}{0.96,0.64,0.38}
\definecolor{seagreen}{rgb}{0.18,0.55,0.34}
\definecolor{seashell1}{rgb}{1.00,0.96,0.93}
\definecolor{seashell2}{rgb}{0.93,0.90,0.87}
\definecolor{seashell3}{rgb}{0.80,0.77,0.75}
\definecolor{seashell4}{rgb}{0.55,0.53,0.51}
\definecolor{seashell}{rgb}{1.00,0.96,0.93}
\definecolor{sienna1}{rgb}{1.00,0.51,0.28}
\definecolor{sienna2}{rgb}{0.93,0.47,0.26}
\definecolor{sienna3}{rgb}{0.80,0.41,0.22}
\definecolor{sienna4}{rgb}{0.55,0.28,0.15}
\definecolor{sienna}{rgb}{0.63,0.32,0.18}
\definecolor{skyblue}{rgb}{0.53,0.81,0.92}
\definecolor{slateblue}{rgb}{0.42,0.35,0.80}
\definecolor{slategray}{rgb}{0.44,0.50,0.56}
\definecolor{slategrey}{rgb}{0.44,0.50,0.56}
\definecolor{snow1}{rgb}{1.00,0.98,0.98}
\definecolor{snow2}{rgb}{0.93,0.91,0.91}
\definecolor{snow3}{rgb}{0.80,0.79,0.79}
\definecolor{snow4}{rgb}{0.55,0.54,0.54}
\definecolor{snow}{rgb}{1.00,0.98,0.98}
\definecolor{springgreen}{rgb}{0.00,1.00,0.50}
\definecolor{steelblue}{rgb}{0.27,0.51,0.71}
\definecolor{tan1}{rgb}{1.00,0.65,0.31}
\definecolor{tan2}{rgb}{0.93,0.60,0.29}
\definecolor{tan3}{rgb}{0.80,0.52,0.25}
\definecolor{tan4}{rgb}{0.55,0.35,0.17}
\definecolor{tan}{rgb}{0.82,0.71,0.55}
\definecolor{thistle1}{rgb}{1.00,0.88,1.00}
\definecolor{thistle2}{rgb}{0.93,0.82,0.93}
\definecolor{thistle3}{rgb}{0.80,0.71,0.80}
\definecolor{thistle4}{rgb}{0.55,0.48,0.55}
\definecolor{thistle}{rgb}{0.85,0.75,0.85}
\definecolor{tomato1}{rgb}{1.00,0.39,0.28}
\definecolor{tomato2}{rgb}{0.93,0.36,0.26}
\definecolor{tomato3}{rgb}{0.80,0.31,0.22}
\definecolor{tomato4}{rgb}{0.55,0.21,0.15}
\definecolor{tomato}{rgb}{1.00,0.39,0.28}
\definecolor{turquoise1}{rgb}{0.00,0.96,1.00}
\definecolor{turquoise2}{rgb}{0.00,0.90,0.93}
\definecolor{turquoise3}{rgb}{0.00,0.77,0.80}
\definecolor{turquoise4}{rgb}{0.00,0.53,0.55}
\definecolor{turquoise}{rgb}{0.25,0.88,0.82}
\definecolor{violetred}{rgb}{0.82,0.13,0.56}
\definecolor{violet}{rgb}{0.93,0.51,0.93}
\definecolor{wheat1}{rgb}{1.00,0.91,0.73}
\definecolor{wheat2}{rgb}{0.93,0.85,0.68}
\definecolor{wheat3}{rgb}{0.80,0.73,0.59}
\definecolor{wheat4}{rgb}{0.55,0.49,0.40}
\definecolor{wheat}{rgb}{0.96,0.87,0.70}
\definecolor{whitesmoke}{rgb}{0.96,0.96,0.96}
\definecolor{white}{rgb}{1.00,1.00,1.00}
\definecolor{yellow1}{rgb}{1.00,1.00,0.00}
\definecolor{yellow2}{rgb}{0.93,0.93,0.00}
\definecolor{yellow3}{rgb}{0.80,0.80,0.00}
\definecolor{yellow4}{rgb}{0.55,0.55,0.00}
\definecolor{yellowgreen}{rgb}{0.60,0.80,0.20}
\definecolor{yellow}{rgb}{1.00,1.00,0.00}

\usepackage{ upgreek }
\usepackage{lscape}
\usepackage{subfigure}
\usepackage{amsmath}
\usepackage{graphicx}
\usepackage{dsfont}
\usepackage{multirow} 
\usepackage{listings,relsize}
\begin{document}
\pagenumbering{Roman} \thispagestyle{empty}
\begin{center} \begin{tabular}{c} \vspace{2in}\\ \begin{minipage}{5in} \begin{center} {\bf \huge
Optimal Patient Allocation in Multi-Arm Clinical Trials\\[1ex]     %your thesis title,
           %note \\[1ex] is a line break in the title

} \vspace{2in}

{\bf \Large Martin Law}

\vspace{2in}

Submitted for the degree of MSc in Statistics\\ at Lancaster University, \\ September 2011.
\end{center} \end{minipage} \end{tabular} \end{center} \newpage

\tableofcontents 
%\listoffigures 
\clearpage 
\pagenumbering{arabic}

\clearpage

\chapter{Preface}

Over the last 20 years, the number of new chemical or biological entities has decreased, despite major increases in pharmaceutical R\&D expenditure. Clinical trials account for 58.6$\%$ of the total cost of R\&D spending, with phase II trials accounting for 15.4$\%$ of the total. The cost of taking a drug to market has also increased, reaching an estimated $\$1,318$ million in 2006 \cite{EFPIA}. Further, 45$\%$ of Phase III trials between 1991-2000 were unsuccessful \cite{kola}. It thus makes sense to look for ways to save money, increase the speed with which new treatments can progress, and perhaps be more cautious when making the decision to proceed to Phase III.\\

% exhibit more care in early phases when doses are being tested.\\
%,and even successful drugs often require a dose change after registration [Cross et al]. 

% yet around of phase 3 trials 50$\%$ fail
As a way to ameliorate the current situation somewhat, Parmar \textit{et al} \cite{parmar} propose multi-arm multi-stage (MAMS) trials: A multi-arm trial is a trial for which a number of active treatment arms are compared to a single control group. This allows the efficacy of a number of treatments to be evaluated more quickly and using less patients than separate single-arm trials. The active treatment arms may be different doses of the same drug, a number of different drugs, or different combinations of drugs. A MAMS trial is a multi-arm trial which includes interim analyses - analysing the data at certain specified points, generally discontinuing treatments which are concluded to not work and proceeding with the remainder. The purpose of such phase II trials is to find the most efficacious treatment (or treatments) from a range of treatments, and proceed to a phase III trial as cost-effectively as possible. Given the expense of conducting clinical trials, any reduction in sample size (and thus cost) is to be welcomed. \\

MAMS trials are now being used in practice, and examples of this include STAMPEDE\cite{stampede}, FOCUS 3 \cite{focus} and ICON6 \cite{icon6}, conducted by the Medical Research Council (MRC). \\

It is possible that the advantages of multi-arm trials over single-arm trials may be enhanced further, by considering the allocation of patients between the control treatment and the active treatments: When designing a multi-arm trial, the proportion of patients allocated to a given treatment is defined by the allocation ratio, $R$. For an $R:1$ allocation ratio, $Rn$ patients are allocated  to the control arm and $n$ patients allocated to each active treatment arm. The definition of an ``optimal'' allocation ratio for a given trial is not fixed; it may be considered optimal to minimise variance, for example. In this study, the optimal allocation ratio will be defined as the allocation ratio which results in the smallest total sample size satisfying some required power and probability of type I error. This is an intuitive definition in the context of clinical trials, as a smaller trial will in general be more ethical (see Discussion) and less expensive than a larger one satisfying the same error rates. Thus, it is of interest to investigate whether total sample size can be considerably reduced through choice of allocation ratio. It would also be non-trivial to find if error rates and sample size can be simply maintained while allocation ratio is varied, as it may be desirable to select the allocation ratio based on the characteristics of one treatment compared with another.\\

For clinical trials with one active treatment arm and one control arm, the optimal allocation ratio can easily be shown to be $1:1$. In the case of multiple active treatment arms, the optimal allocation ratio has not been explicitly investigated, and to do so is the purpose of this study.\\

\begin{figure}[htbp]
	\centering
		\includegraphics[width=0.5\textwidth]{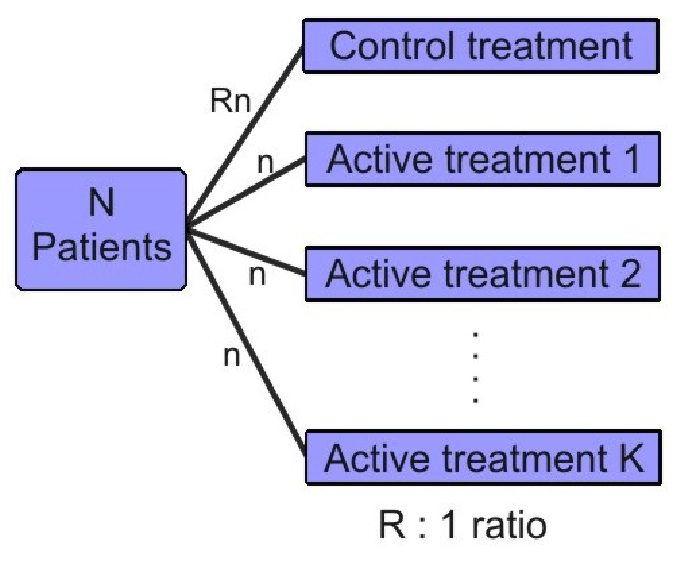}
	\caption{Clinical trial with K active treatment arms and allocation ratio R:1}
	\label{fig:MAmodel9}
\end{figure}

\begin{figure}[htbp]
	\centering
		\includegraphics[width=0.5\textwidth]{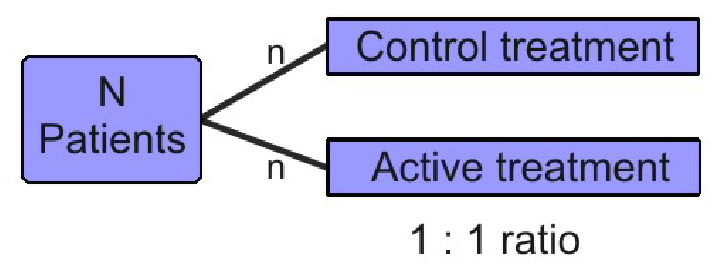}
	\caption{Two-arm clinical trial with allocation ratio 1:1}
	\label{fig:2Amodel9}
\end{figure}

The structure of the main body of this study is as follows: The setup for a single stage trial with $K$ active treatment arms is described in Section 2, along with a brief exposition of Dunnett's statement regarding the optimal allocation ratio in such circumstances \cite{dunnett} . Equations for type I error and power are derived, and the methodology used to investigate how total sample size may be minimised using allocation ratio is described. A two-stage trial is then considered, using the same methodology. Figures and tables showing how total sample size changes with allocation ratio, for a range of type I error and power values, are given in Section 3. The possible ethical and financial benefits of changing allocation ratio, including a simple example, is also included in Section 3. The results, and what they could mean in practical terms, are discussed in Section 4.\\

% Finally, the inclusion of an interim analysis is then considered, which allows early stopping of the trial for efficacy or futility.\\
% giving a total sample size of $N = Rn + Kn$.

\clearpage

\chapter{Methods}
\subsection{One-stage design}

Consider a clinical trial with one control arm and $K$ active treatment arms, with an $R:1$ allocation ratio - that is, $Rn$ patients allocated to the control arm and $n$ patients allocated to each active treatment arm - giving a total sample size of $N = Rn + Kn$. For each active treatment $i$, a test statistic $Z_i$ is calculated by comparing that treatment's results to those of the control treatment. The greatest of these test statistics, $Z_M$, then represents the active treatment which has performed best. If $Z_M > C$  for some critical value $C$, we conclude that this treatment has an efficacy worthy of further investigation, and we may proceed to the next phase. Otherwise, we conclude that the treatment does not have an efficacy worth pursuing, and the treatment is abandoned. In both cases, all other treatments are not considered further.\\

Define trial results for treatment $i$ as

	\begin{equation}
X_{ij} \sim N(\mu_i,\sigma^2), i = 0,1,2,\dotsc, K. \, \  j = 1, 2,\dotsc, n, \nonumber
%\label{eq:}
\end{equation}

The distribution of the mean of results for the control treatment, $\bar{X_0}$, is
\begin{eqnarray}
\bar{X_0} & \sim & N \left(\mu_0, \frac{\sigma^2}{Rn}\right). \nonumber
%\label{eq:xbar}
\end{eqnarray}

The distribution of the mean of results for treatment $i$ is
\begin{eqnarray}
\bar{X_i} & \sim & N \left(\mu_i, \frac{\sigma^2}{n}\right),\; i = 1, 2,\dotsc , K. \nonumber
%\label{eq:xbar}
\end{eqnarray}

With $\bar{X_i}$ as above, the distribution of the sum of the outcomes, $S_i = \sum_{j=1}^n X_{ij},  $  is $S_i \sim N(n\mu_i, n\sigma^2), \; i = 1, 2,\dotsc , K, $ for the active treatments and $S_0 \sim N(Rn\mu_0, Rn\sigma^2)$ for the control treatment.\\

It is wished to test the null hypothesis that all active treatments are no better than the control: $H_0:= \mu_0 = \mu_1 = \dots = \mu_K = 0$, where $\mu_i$ is the true mean effect of the treatment $i$. The effect of treatment $i$ is compared to the control using a standard one-sided $Z$-test, using the following test statistic:\\

\begin{eqnarray}
Z_i & =  &  \frac{\bar{X_i} - \bar{X_0}}{\sigma \sqrt{\frac{1}{n}+\frac{1}{Rn}}}  \nonumber  \\
		& =     &  \frac{\frac{S_i}{n} - \frac{S_0}{Rn}}{\sigma\sqrt{\frac{R+1}{Rn}}}\nonumber\\
		& =			&	 \frac{RS_i - S_0}{\sigma  \sqrt{(R+1)Rn}}
\label{eq:TSS}
\end{eqnarray}

%%%%%%%%%%%%%%%%%%%%%%%%%%%%%%%%%%%%%%%%%
% WHAT DUNNETT DID
%%%%%%%%%%%%%%%%%%%%%%%%%%%%%%%%%%%%%%%%%

\subsection{Dunnett's result}

Dunnett \cite{dunnett} states, without derivation, that the optimal allocation ratio in a multi-arm trial as defined above is $\sqrt{K}:1$. One can show this, under some simplification, in the following scenario:\\

%Assume that there is only one active treatment. 
Relax the requirement that a test statistic $Z_i$ must be the maximum to be considered an efficacious treatment. Then the power of the trial can be defined as

\begin{equation}
P(Z_i > C \;| \; H_A:= \text{Treatment $i$ works satisfactorily}) = P\left(\frac{\bar{X_i}-\bar{X_0}}{\sigma\sqrt{\frac{1}{n}+\frac{1}{Rn}}} > C \middle\vert H_A \right). \nonumber
\label{eq:dunn}
\end{equation}

The optimal allocation ratio may be considered that which will maximise the power. To maximise the power, we must minimise $(1/n) + (1/Rn)$, or equivalently, minimise $\frac{R+K}{N} + \frac{R+K}{RN}$ (using the substitution $N = Rn + Kn$). Differentiating and setting equal to 0 gives:

\begin{eqnarray}
\frac{d}{dR} \left( \frac{R+K}{N} + \frac{R+K}{RN} \right) & = & 0 \nonumber \\
%\Rightarrow \frac{d}{dR} \left( \frac{R(R+K) + R + K}{RN}\right) & = & 0 \nonumber \\
\Rightarrow \frac{d}{dR} \left( \frac{R}{N} +  \frac{K}{N} + \frac{1}{N} + \frac{K}{RN}\right) & = & 0 \nonumber \\
\Rightarrow \frac{1}{N} - \frac{K}{R^2N} & = & 0 \nonumber \\
\Rightarrow \frac{1}{N} & = & \frac{K}{R^2N}\nonumber \\
%\Rightarrow R^2 & = & \frac{KN}{N}\nonumber \\
\Rightarrow R & = & \sqrt{K}\nonumber 			 
\end{eqnarray}

Hence it may be argued from a theoretical perspective that the allocation ratio $\sqrt{K}:1$ is, if not optimal exactly, at least a reasonable approximation. It may be regarded as an approximation due to the simplification employed in (\ref{eq:dunn}) - that is, relaxing the requirement that the test statistic calculated for treatment $i$ is the maximum of all test statistics, and focusing only on the requirement that it must be greater than the critical value $C$, given that the treatment works satisfactorily.\\

\subsection{Type I error}

In trials with one active treatment arm, a type I error is made when it is concluded to take the active treatment forward when its true treatment effect is not clinically relevant.
%below the minimum clinically relevant difference. 
For multi-armed trials, a type I error is made when any active treatment is taken forward erroneously. Recall that the decision to take a treatment forward is made if the treatment performs the best - that is, its test statistic is the greatest of all test statistics - and the test statistic for that treatment exceeds some critical value $C$. Therefore, a type I error is made if the greatest test statistic $Z_M$ exceeds critical value $C$ given that the true treatment effect is not clinically relevant. Thus the probability of a type I error occurring, $\alpha$, is

%			P \left(    \right) 

\begin{eqnarray}
\alpha  & = & P (\text{take any active treatment forward}| H_0:= \mu_0 = \mu_1 = \dots = \mu_K = 0)\nonumber \\
				& = & P \left(  Z_M \geq C \:|\: H_0   \right)  \nonumber \\
				%\text{, where $Z_M$ is the maximum of all test statistics} \nonumber \\
\Rightarrow 1 - \alpha  & = & P \left(  Z_M < C \:|\: H_0   \right). \nonumber \\
 & = & P \left( Z_1 < C ,\: Z_2 < C , \dotsc , Z_K < C \:|\: H_0 \right) \nonumber 
 %\label{eq:}
\end{eqnarray}

$1-\alpha$ is shown in Appendix A to be equal to

\begin{eqnarray}
1 - \alpha & = & \int_{-\infty}^\infty \! \left[ 	\Phi \left(	C	\sqrt{\frac{R+1}{R}} + \frac{x}{\sqrt{R}} 	\right)		\right]^K \phi (x) \; dx,  
\label{eq:11}
\end{eqnarray}

%where $x = \frac{s - Rn\mu}{\sqrt{Rn}\sigma}$.\\

%%%%%%%%%%%%%%%%%%%%%%%%%%%%%%%%%%%%%%%%%
% TYPE II EQUATION
%%%%%%%%%%%%%%%%%%%%%%%%%%%%%%%%%%%%%%%%%

\subsection{Power}

%In the multi-arm case, the power of the study may be defined in a number of ways [ optimal design 2007]. Here, the ``minimal power'' definition will be used. This refers to the probability of rejecting \textit{at least one} statement of the null hypothesis that is false.\\

Consider $\delta$, a treatment effect of magnitude such that we would like to take forward any treatment for which $\mu \geq \delta$. Also consider $\delta_0$, a treatment effect of magnitude such that we would not like to take forward any treatment for which $\mu < \delta_0$. Note that $0 \leq \delta_0 < \delta$, and for $\delta_0 < \mu < \delta$, both taking the treatment forward and dropping the treatment are deemed acceptable. Without loss of generality, assume that under the alternative hypothesis, the true effect of treatment $K$ is $\delta$. The power of the study - the probability of taking forward a treatment clinically relevant effect, is then defined as

\begin{eqnarray}
1 - \beta & = & P(\text{Take treatment $K$ forward} | H_A:= \mu_0 = 0, \mu_1 = \mu_2 = \dots = \mu_{K-1} = \delta_0, \mu_K = \delta) \nonumber \\
					& = & P (Z_K = Z_M,\: Z_K \geq C \:|\: H_A) 
\label{eq:powZ}
\end{eqnarray}

This is known as the \textit{least favourable configuration} \cite{thall}, and has been employed in related studies \cite{johnjack, stall}. This hypothesis - where a single treatment has a clinically relevant effect, while all others have the greatest effect which is not clinically relevant - makes it difficult to identify the one treatment of interest. Indeed, this hypothesis minimises power by being effectively a ``worst case scenario''; the actual power of the trial will exceed the theoretical power in any other circumstance. The motivation for using this hypothesis is to ensure that the power specified is always reached.\\

%Power increases as $\delta$ increases or $\delta_0$ decreases, and so the actual power of the study should be greater than or equal to the value specified by this definition. The motivation for using this definition of power is to ensure that the power specified is always reached.\\

The power (\ref{eq:powZ}) is shown in Appendix B to be

\begin{eqnarray}
1 - \beta		& = & \int_{-\infty}^{\infty} \left[  \Phi  \left(	 w + \frac{\sqrt{n}}{\sigma} (\delta - \delta_0)	\right) \right]^{K-1} \Phi \left( w \sqrt{R} + \frac{\sqrt{Rn}\delta}{\sigma} - C \sqrt{R+1} \right) \phi(w) dw. \nonumber\\
\label{eq:t22}
\end{eqnarray}

\subsection{Simulation}

Equations (\ref{eq:11}) and (\ref{eq:t22}) were verified using simulation: For differing configurations of $\alpha,$ power, $  K, R, \sigma, \delta \text{ and } \delta_0$, 10,000 trials were simulated, and the resulting test statistics for each active treatment calculated. The probability of a type I error occurring was found by simulating data such that the null hypothesis ($H_0:= \mu_0 = \mu_1 = \dots = \mu_K = 0$) was true, and finding the proportion of times the best-performing treatment was concluded to have a clinically relevant effect. The power was found by simulating data such that the alternative hypothesis ($H_A:= \mu_0 = 0$, $\mu_1 = \mu_2 = \dots = \mu_{K-1} = \delta_0$, $\mu_K = \delta$) was true, and finding the proportion of times treatment $K$ was both the best-performing treatment and concluded to have a clinically relevant effect.\\

%that the treatment effect of treatment $K$ was $\delta$, the treatment effect of treatments $1, 2, \dotsc , K-1 $ was $\delta_0$, and the treatment effect of the control treatment was $0$, then finding the proportion of time treatment $K$ would have been taken forward. \\

\section{Function \textit{find.all}}

It can be seen from (\ref{eq:11}) that the probability of a type I error occurring, $\alpha$, is a function of the number of active treatment arms $K$, the allocation ratio $R$ and the critical value $C$, but does not depend on $n, \sigma, \delta$ or $\delta_0$. Thus the smallest critical value satisfying a required $\alpha$ can be found by specifying merely $K$ and $R$. With $K, R, \alpha \text{ and } C$ known, the smallest total sample size, $N$, satisfying a required power (and $\alpha$) can then be found using (\ref{eq:t22}) by specifying the remaining values - $\sigma, \delta,\text{ and } \delta_0$.\\

Under this premise, a function, \textit{find.all}, was created using the statistical package R 2.11.1 \cite{R}: Given requirements for $\alpha$ and power, and values for $R, K, \sigma, \delta$ and $\delta_0$, \textit{find.all} finds the smallest total sample size (and associated critical value) which satisfies the given type I error and power requirements. This total sample size is returned, as well as the change in total sample size in comparison to using a standard $1:1$ allocation ratio, both in number of patients and as a proportion. The function performs this for a range of allocation ratios, with the user specifying both the range and increment size.  For non-integer values of $R$, $Rn$ and thus $N$ may too be non-integer values; this would not make sense in the context of clinical trials. To arrive at integer values, the function simply rounds up $Rn$ to the nearest integer, resulting in an allocation ratio as close to $R:1$ as possible. The function also evaluates the reduction in sample size of each active treatment arm. Finally, a plot is returned, giving a visual indication of how sample size changes with allocation ratio. The code for \textit{find.all} is included in full in Appendix C.\\

%A function was created to automate this procedure and plot how $N$ varies with $R$, using statistical package R - the code is included in the Appendix. \\

A number of sets of arguments were used in order to investigate how sample size varies with allocation ratio. All argument values used are listed below, with the ``default'' values in bold:

\begin{itemize}
	\item $\alpha = $ 0.025, \textbf{0.05}, 0.1, 0.2;
	\item Power = 0.8, \textbf{0.9};
	\item $\sigma =$ \textbf{1}, 1.25, 1.5, 1.75;
	\item $\delta =$ \textbf{0.5}, 0.6, 0.7, 0.8;
	\item $\delta_0 =$ 0.075, \textbf{0.125}, 0.175, 0.225.
\end{itemize}

One argument was varied while the rest remained fixed. The default values for $\alpha$ (0.05) and power (0.9) are common values for trials. The default value for $\delta (0.5)$ was chosen to be one half of the standard deviation $\sigma (1)$, while the default value for $\delta_0 (0.0125)$ was chosen to be one quarter of $\delta$ - a proportion used previously \cite{johnjack}. Total sample size was plotted against allocation ratio $R$, for ratios from 1:1 to 5:1, and in increments of size 0.1. Such plots were created for a range of active treatments: $K = 2, 3, 4, 5$, and are the results for varying $\alpha$ and power are shown in Figures \ref{fig:a} and \ref{fig:beta} in Section 3, and for varying $\sigma, \delta$ and $\delta_0$ in Appendix C.\\

With these results, it was hoped to be able to answer the following questions:\\
\begin{itemize}
	\item Is the true optimal allocation ratio $\sqrt{K}:1$ as suggested by Dunnett?
	\item Does choosing the optimal allocation ratio result in a considerable reduction in sample size?
	\item Are there any other benefits to changing allocation ratio?
\end{itemize}

The second and third questions effectively seek to answer the same underlying question:\\

\begin{center}
	\textit{Can clinical trials be improved through our choice of allocation ratio?}
\end{center}

\newpage

\section{Two-stage design}

It could be argued that, as only one active treatment at most will be concluded to work to a satisfactory degree, having $K$ active treatment arms for the full duration of the clinical trial is not an optimal design. Indeed, it is possible to design a trial in which the effect of the $K$ active treatment arms are evaluated at an interim stage, following which the trial continues with only one active treatment alongside the control.\\

Formally, consider the addition of an interim analysis to the one-stage design. At this interim, results are available for $Rn_1$ patients on the control treatment and $n_1$ patients for each active treatment arm. Again, a test statistic is found for each active treatment. Interest is in the best-performing treatment, that with the greatest test statistic $Z_M$. At this interim, we \emph{select} the best-performing treatment, dropping all others, and the trial continues with only two treatment arms. After the interim, $Rn_2$ patients are allocated to the control treatment and $n_2$ to the remaining active treatment. At the end of the second stage, another test statistic, $Z_{2*}$, is calculated, taking account of the new data. If $Z_{2*} \geq C$ for some critical value $C$, then we conclude that the treatment works. If not, we conclude that the treatment does not work. As before, the other treatments are ignored. We fix $n_1 = n_2 = n$, to ensure that of the patients receiving the control and best-performing active treatments, half do so before the interim, and half after. The total sample size of the trial is then $N=Rn + Kn + Rn + n = n(2R + K + 1)$. Define the test statistic at the second stage as

\begin{eqnarray}
Z_{2*} & = & \frac{\frac{S_{2*}}{2n} - \frac{S_{20}}{2Rn}}{\sigma \sqrt{\frac{1}{2n} + \frac{1}{2Rn}}} \nonumber \\
       & = & \frac{RS_{2*} - S_{20}}{\sigma \sqrt{(R+1)2Rn}}\nonumber \\
       & = & \frac{R(S_{1M}+S_+) - (S_{10}+S_{+0})}{\sigma \sqrt{(R+1)2Rn}}\nonumber \\
       & = & \frac{RS_{1M}-S_{10}}{\sigma \sqrt{(R+1)2Rn}} + \frac{RS_+-S_{10}}{\sigma \sqrt{(R+1)2Rn}} \nonumber \\
       & = & Z_{1M} \frac{1}{\sqrt{2}} + \frac{RS_+-S_{10}}{\sigma \sqrt{(R+1)2Rn}}, \nonumber 
\label{eq:}
\end{eqnarray}

where $Z_{1M}$ is the test statistic of the best-performing treatment. The probability of a type I error occurring is

\begin{eqnarray}
\alpha  & = & P (\text{take any active treatment forward}| H_0:= \mu_0 = \mu_1 = \dots = \mu_K = 0)\nonumber \\
	   & = & P \left(  Z_{2*} \geq C \:|\: H_0   \right),  \nonumber
%\label{eq:}
\end{eqnarray}

while the power of the study is

\begin{eqnarray}
1 - \beta & = & P(\text{Take treatment $K$ forward} | H_A:= \mu_0 = 0, \mu_1 = \mu_2 = \dots = \mu_{K-1} = \delta_0, \mu_K = \delta) \nonumber \\
		& = & P (Z_K = Z_M,\: Z_{2*} \geq C \:|\: H_A) \nonumber 
%\label{eq:powZ}
\end{eqnarray}

The type I error and power equations have been derived previously \cite{johnjack}, for the case where there are three active treatment arms and the allocation ratio is $1:1$. In the previous derivation, $r = \frac{n_1 + n_2}{n_1}$, which is equal to 2 here, as $n_1 = n_2 = n$.  What follows are the equations, generalised to allow allocation ratio $R$ to vary:\\

\begin{eqnarray}
1 - \alpha 	& = & \int_{-\infty}^{\infty} \int_{-\infty}^{\infty} \left[  \Phi  \left(	 A + C\sqrt{\frac{2(R+1)}{R}} - B\sqrt{\frac{R+1}{R}} \right) \right]^K \phi(A) \phi(B) \: dA\, dB
\label{eq:t1.select}\nonumber \\
& & \\
1 - \beta	  & = & \int_{-\infty}^{\infty} \int_{-\infty}^{\infty}  \Phi \left( w - C\sqrt{\frac{2(R+1)}{R}} + u \sqrt{\frac{R+1}{R}} + 2\sqrt{n}\frac{\delta}{\sigma} \right)\nonumber\\
		  &   & \times \left[  \Phi  \left(	 w + \frac{\sqrt{n} (\delta - \delta_0)}{\sigma}	\right) \right]^{K-1} \phi(w) \phi(u) \: dw\, du. 
\label{eq:p.select}
\end{eqnarray}

The equations (\ref{eq:t1.select}) and (\ref{eq:p.select}) were verified using simulation under the null and alternative hypotheses, in the same manner as for the equations in the single-stage case. Again as for the single-stage case, the smallest critical value satisfying a required $\alpha$ may be found by specifying simply the number of treatment arms $K$ and the allocation ratio $R$, using (\ref{eq:t1.select}). The smallest total sample size $N$ satisfying a required power and $\alpha$ may then be found using these values and specifying $\sigma, \delta$ and $\delta_0$ for (\ref{eq:p.select}). \\

\subsection{Function \textit{two.stage}}

The function used for the single-stage design, \textit{find.all}, was altered to use the two-stage equations (\ref{eq:t1.select}) and (\ref{eq:p.select}). However, the method used by the function - finding $\alpha$ for a range of (3,000) critical values, choosing the most suitable one, then using this to find the power for a range of total sample sizes $N$ - was extremely slow. This was due to the equations being double integrals rather than single, and would have taken approximately one month for the desired results to be calculated.\\

To solve this problem, a new function was created, \textit{two.stage}. This function uses the same principle of finding the smallest suitable critical value, then using this to find the smallest total sample size, but does so using the bisection method: Maximum and minimum values of the critical value $C$ are set, and $\alpha$ is found for the midpoint of these. The midpoint becomes either the new maximum or minimum, depending on whether the calculated value of $\alpha$ is greater or less than the required type I error, and this process is iterated until $C$ is found to a high degree of accuracy. A similar process takes place for finding the total sample size $N$. Some of the code used was taken from a function already created to undertake a similar task \cite{amy}.\\

This function was used to investigate how total sample size changes with allocation ratio in a similar manner as for the single-stage design: The same ``default'' set of arguments was used, and one varied while the rest remained fixed. Total sample size was plotted against allocation ratio, for ratios from $1:1$ to $5:1$, again in increments of $0.1$, and for the same range of active treatments: $K = 2, 3, 4, 5$. As with the single-stage case, the underlying question is: Can clinical trials be improved through our choice of allocation ratio?

%For fixed sets of arguments, the minimum possible sample size N was compared across a range of allocation ratios, and the results noted. The effect of changing other argument values - $\alpha, \beta, \sigma, \delta, \delta_0, (\delta-\delta_0)$ - was noted.
		
\clearpage 

\chapter{Results}
\section{Single-stage design}
\subsection{Optimal allocation ratio}

The optimal allocation ratio, $R_{OP}$, for a range of type I error, power and number of active treatments is presented in Table \ref{tab:optimR}. Though $R_{OP}$ does increase as the number of active treatments $K$ increases, it is always $\leq \sqrt{K}$, Dunnett's suggestion of the optimal allocation ratio. This disparity may be attributable to the fact that $\sqrt{K}$ comes from a simplification; in application, there is added complexity, including the choice of argument values for the least favourable configuration. In some instances, $R_{OP}$ is given as a range of values - for example $R_{OP} = 1.5-1.7$. In this case, setting $R = 1.5, 1.6$ or $1.7$ will all give the smallest total sample size, and so any of these values may be considered to be optimal. The optimal allocation ratio tends to not decrease as type I error becomes more strict, but there appears to be no simple pattern explaining the trends of the tabulated values. \\

\begin{table}[htb]
	\centering
		\begin{tabular}{cc|cc| cc| cc| cc}
		
		& & \multicolumn{8}{c}{\textbf{Active treatments} $\mathbf{K (\sqrt{K})}$}\\
		  &  & \multicolumn{2}{c}{\textbf{2 (1.41)}} & \multicolumn{2}{c}{\textbf{3 (1.73)}}& \multicolumn{2}{c}{\textbf{4 (2.00)}}& \multicolumn{2}{c}{\textbf{5 (2.24)}} \\
	     \hline
	     \textbf{Power}& $\mathbf{1 - \boldsymbol\beta}$			& \textbf{0.8} 	& \textbf{0.9}	& \textbf{0.8} 	& \textbf{0.9}	& \textbf{0.8} 	& \textbf{0.9}	& \textbf{0.8} 	& \textbf{0.9}\\
	     \hline
	     &&&&&&&&&\\
	    										& \textbf{0.2}	& 1.0	& 1.2	& 1.5	& 1.3	& 1.6	& 1.3 	& 1.7	& 1.4-1.6\\
	     &&&&&&&&&\\
	      \textbf{Type I}									& \textbf{0.1}	& 1.1-1.3 & 1.2	& 1.2	& 1.4	& 1.8	& 1.5	& 1.9	& 1.5\\
	     \textbf{error}&&&&&&&&&\\ 		
	      		$\boldsymbol\alpha$				& \textbf{0.05}	& 1.4	& 1.2	& 1.5	& 1.6	& 1.6-2.0	& 1.9	& 2.3	& 1.8\\
	     &&&&&&&&&\\		
	      									& \textbf{0.025}	& 1.3	& 1.4	& 1.5-1.7 & 1.5-1.6 & 1.7	& 1.7	& 1.8-2.3	& 2.1\\
		\end{tabular}
	\caption{Optimal allocation ratio $R_{OP}$. }
	\label{tab:optimR}
\end{table}
 %\multirow{7}{*}{$\boldsymbol\alpha$} 

%The difference between $R_{OP}$ and $\sqrt{K}$ tends to increase as $K$ increases, and also as type I error requirement becomes more stict.\\ <<< TOTAL NONSENSE!

\subsection{Reducing total sample size}

Plots of how total sample size changes with allocation ratio for differing type I error and power requirements are shown in Figures \ref{fig:a} and \ref{fig:beta} respectively, with figures for other arguments included in Appendix C. The allocation ratio $\sqrt{K}:1$ is represented by a vertical line. The ``jagged'' nature of the curves is due to rounding to maintain the integer nature of the sample sizes. The curves appear to be quadratic, with minima inside the interval $R_{OP} \in[1, \sqrt{K}]$. \\

\begin{figure}[ht]
  \begin{center}
   \subfigure[Active treatments = 2]{\includegraphics[scale=0.41]{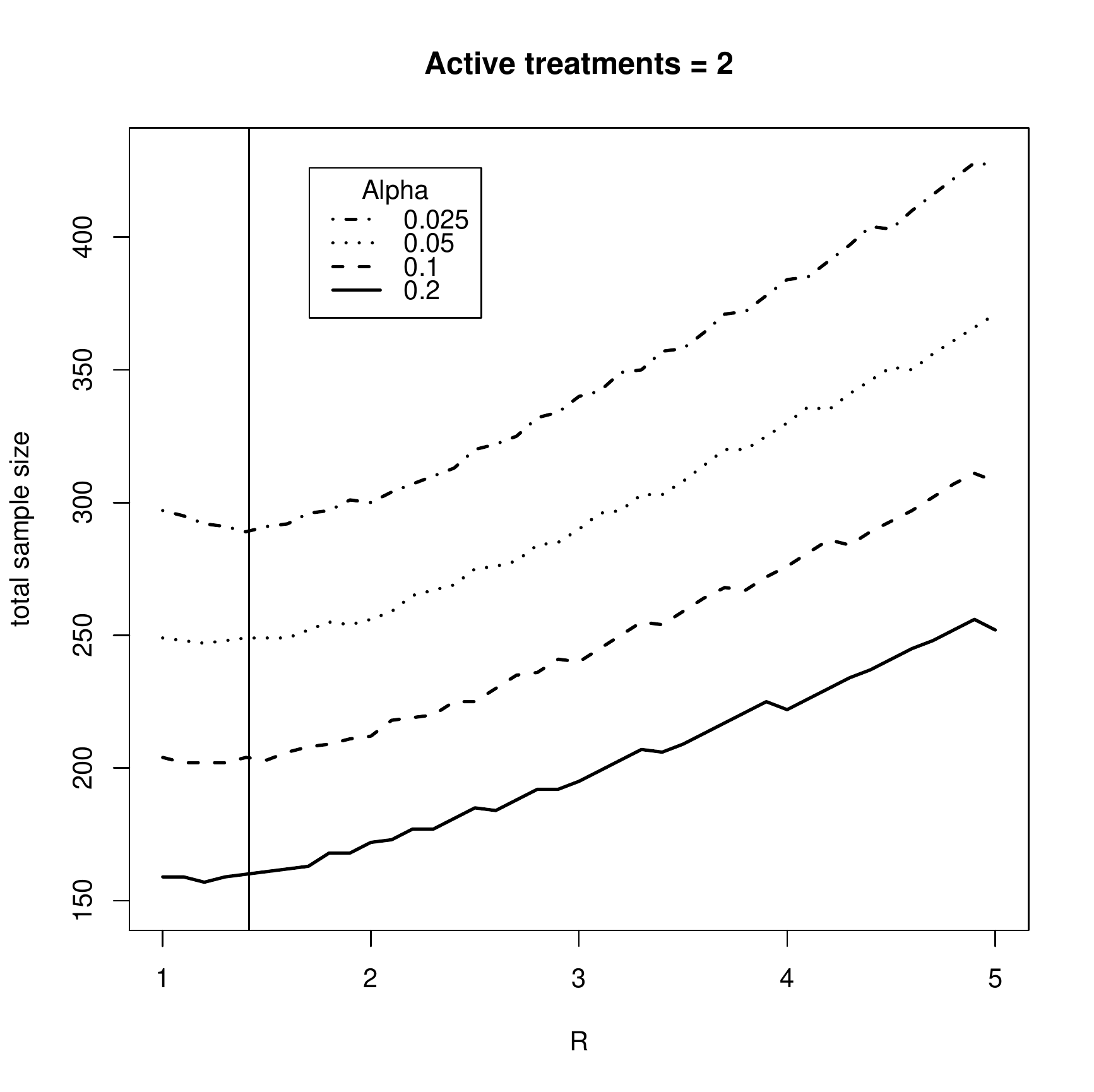}}
    \subfigure[Active treatments = 3]{\includegraphics[scale=0.41]{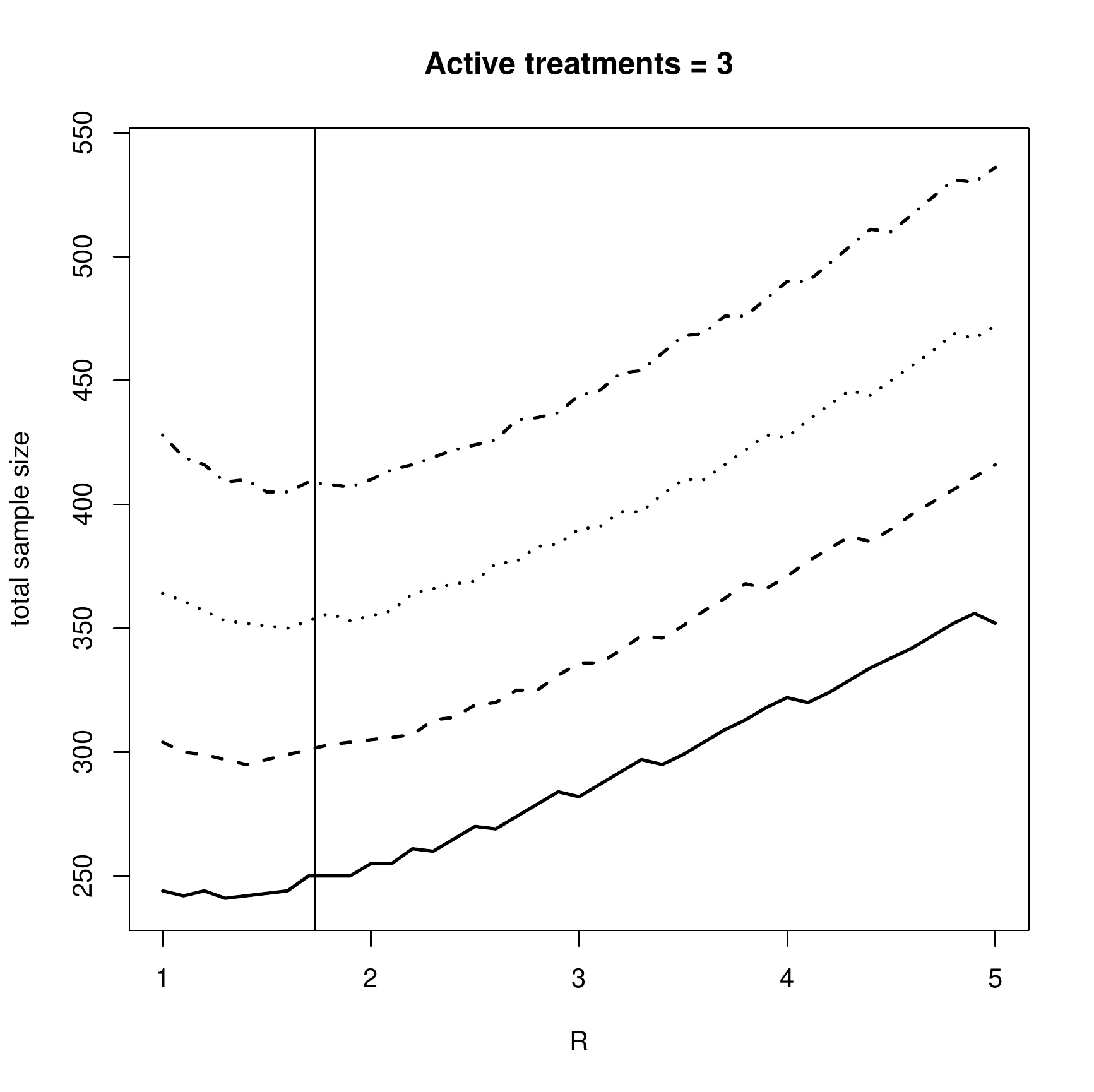}}
     \subfigure[Active treatments = 4]{\includegraphics[scale=0.41]{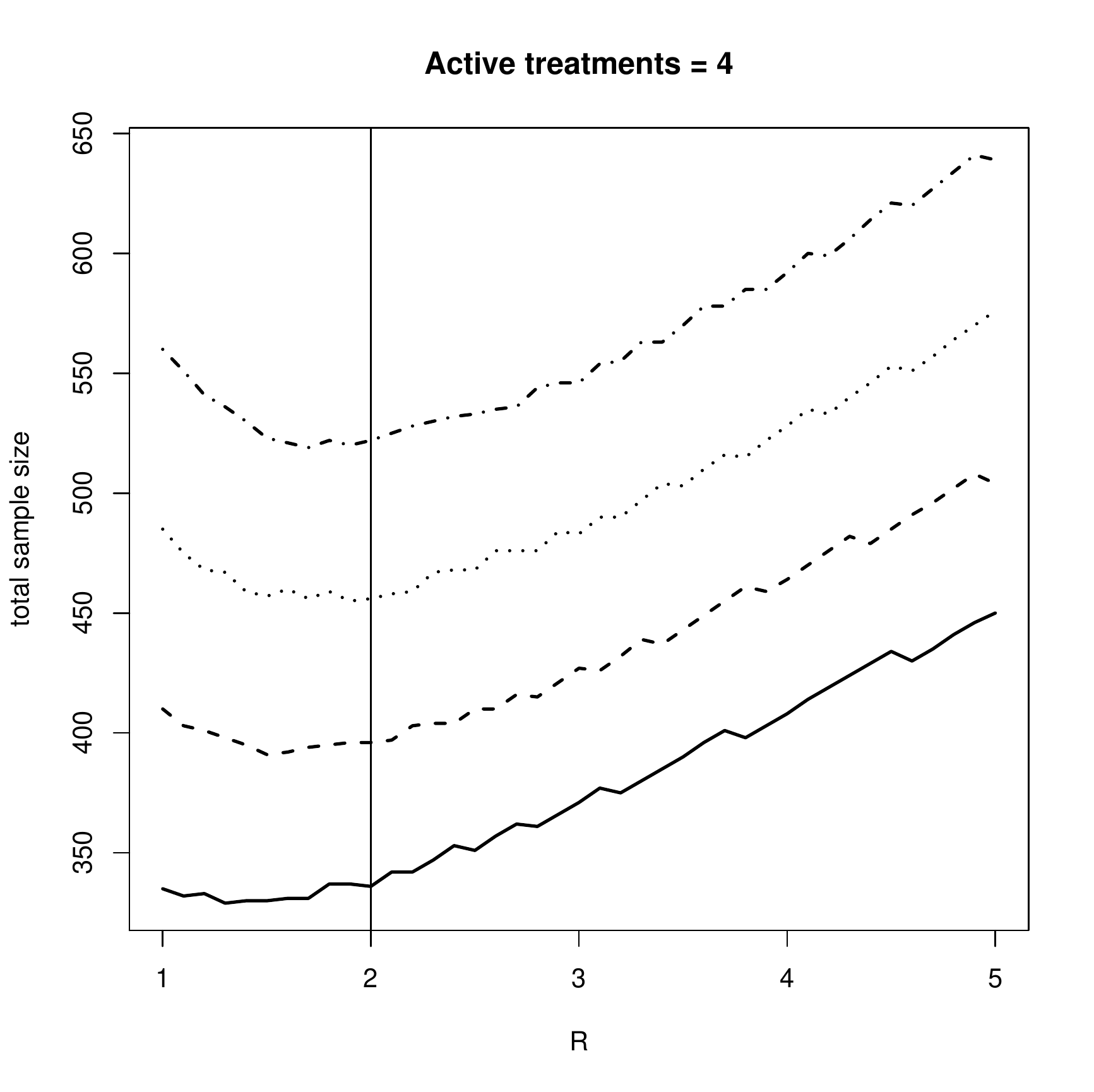}}
      \subfigure[Active treatments = 5]{\includegraphics[scale=0.41]{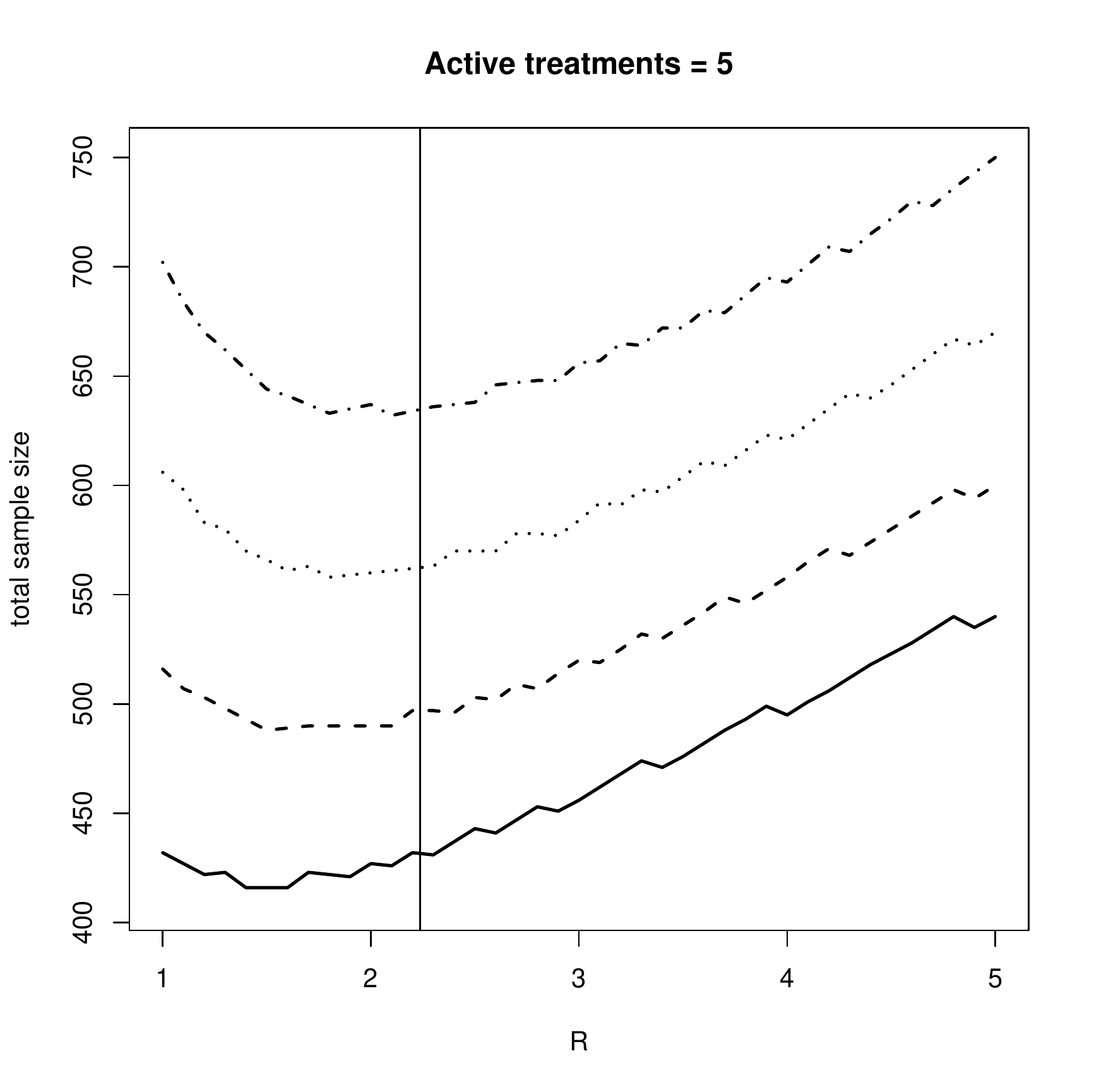}}
  \end{center}
  \caption{Change in total sample size for $\alpha = 0.025, 0.05, 0.1, 0.2$ and $K = 2, 3, 4, 5$, with fixed arguments power $= 0.9, \sigma = 1, \delta = 0.5,$ and $\delta_0 = 0.125$}
  \label{fig:a}
  \end{figure}

\begin{figure}[ht]
  \begin{center}
   \subfigure[Active treatments = 2]{\includegraphics[scale=0.41]{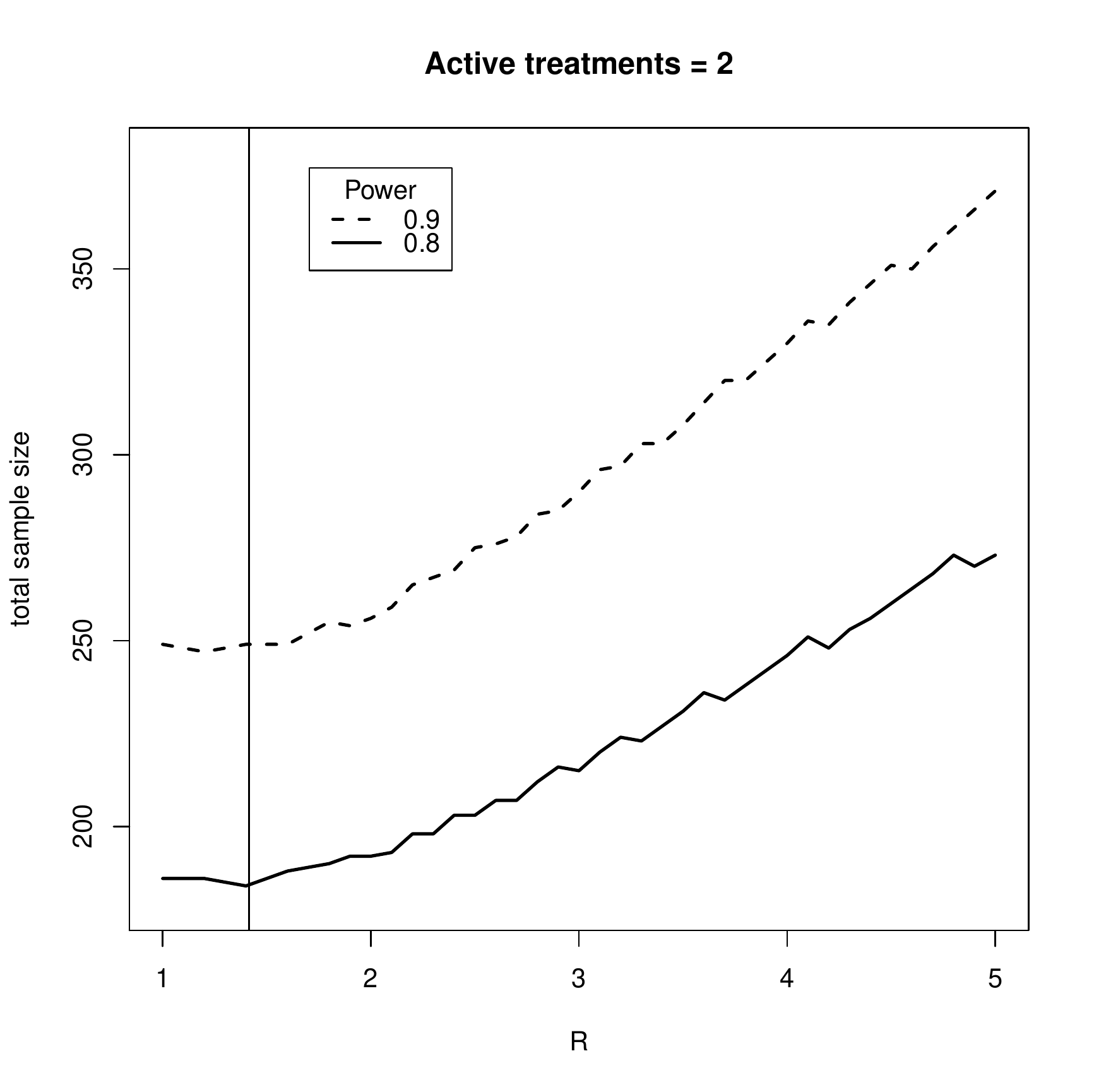}}
    \subfigure[Active treatments = 3]{\includegraphics[scale=0.41]{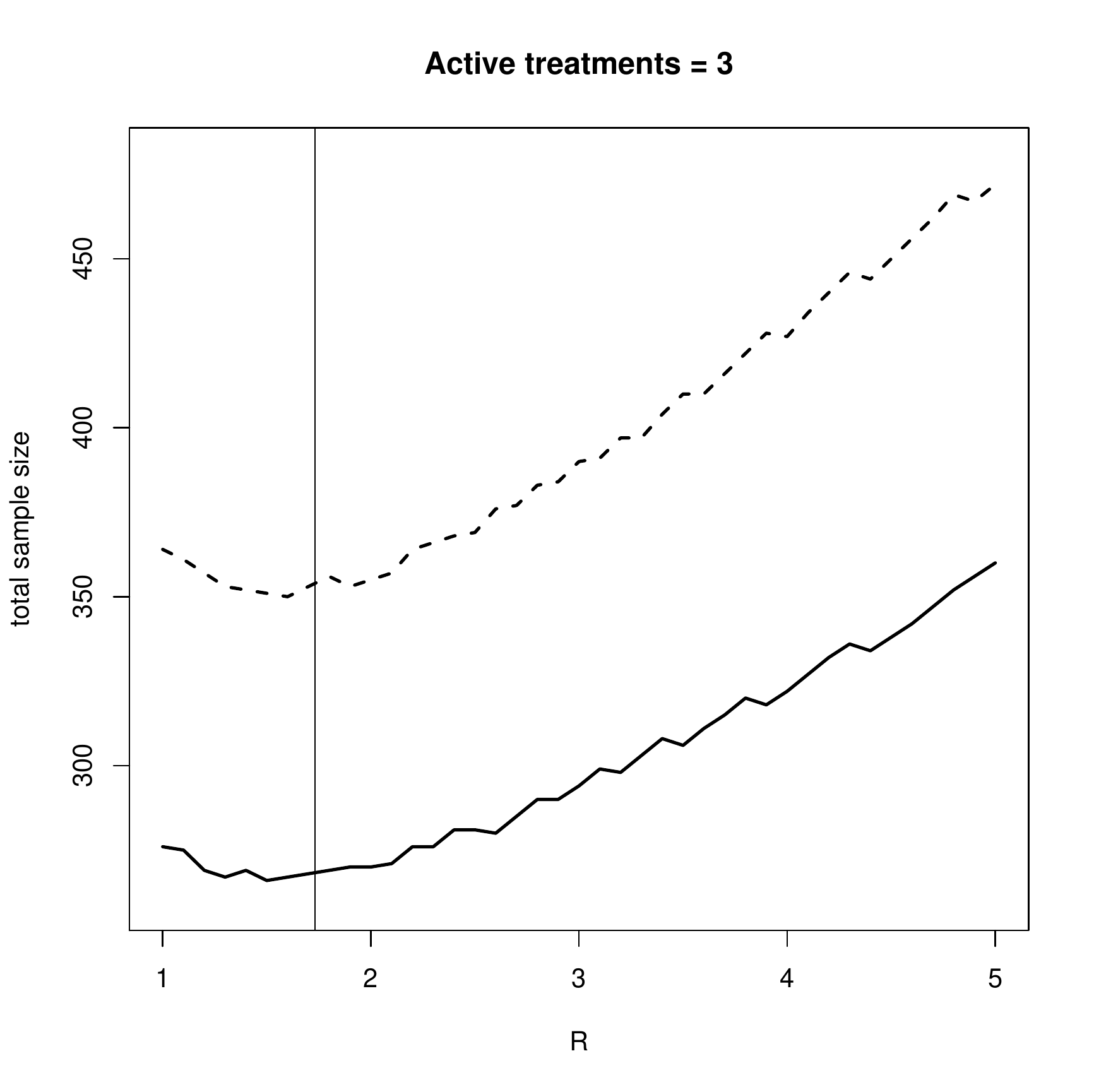}}
     \subfigure[Active treatments = 4]{\includegraphics[scale=0.41]{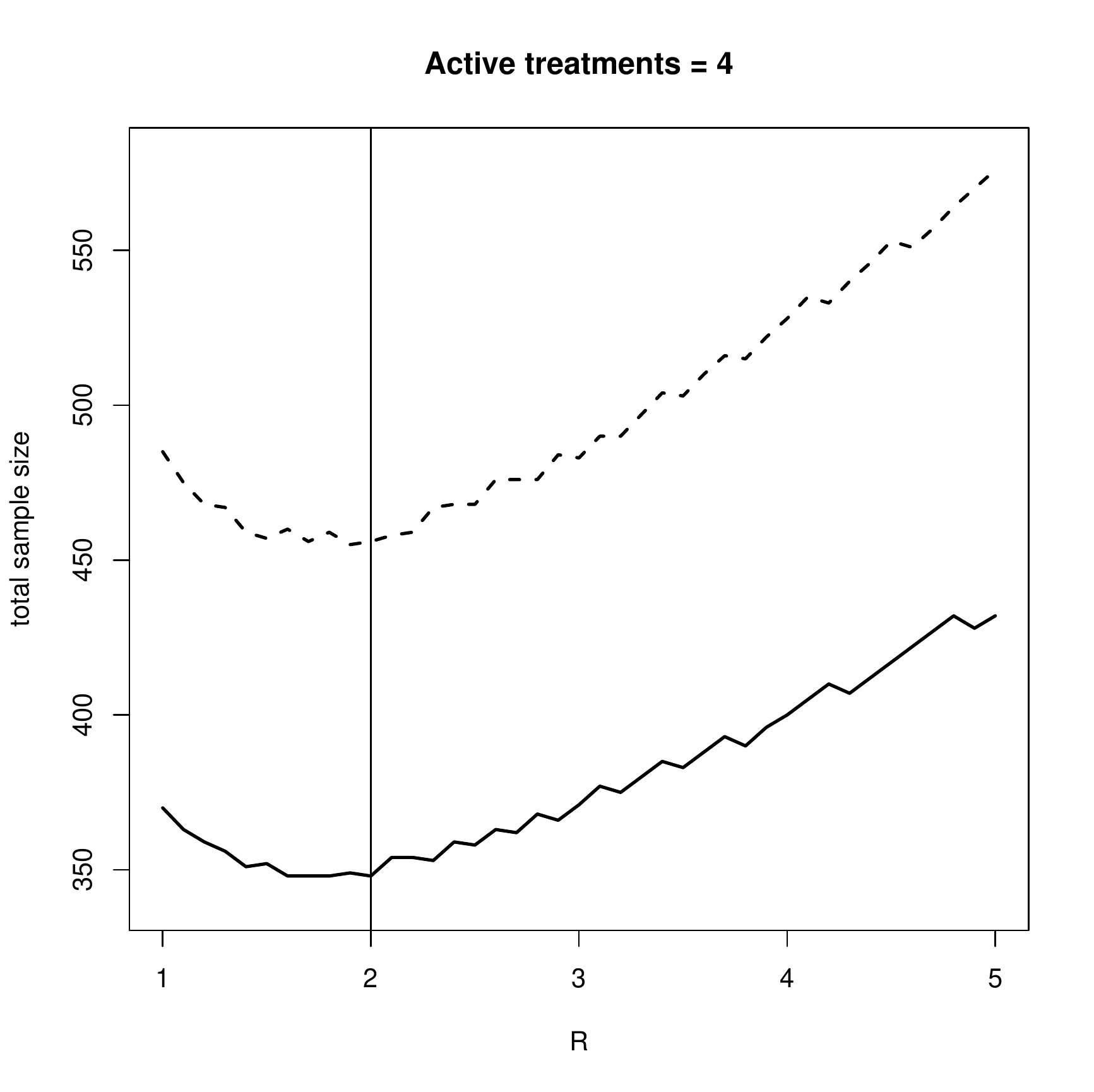}}
      \subfigure[Active treatments = 5]{\includegraphics[scale=0.41]{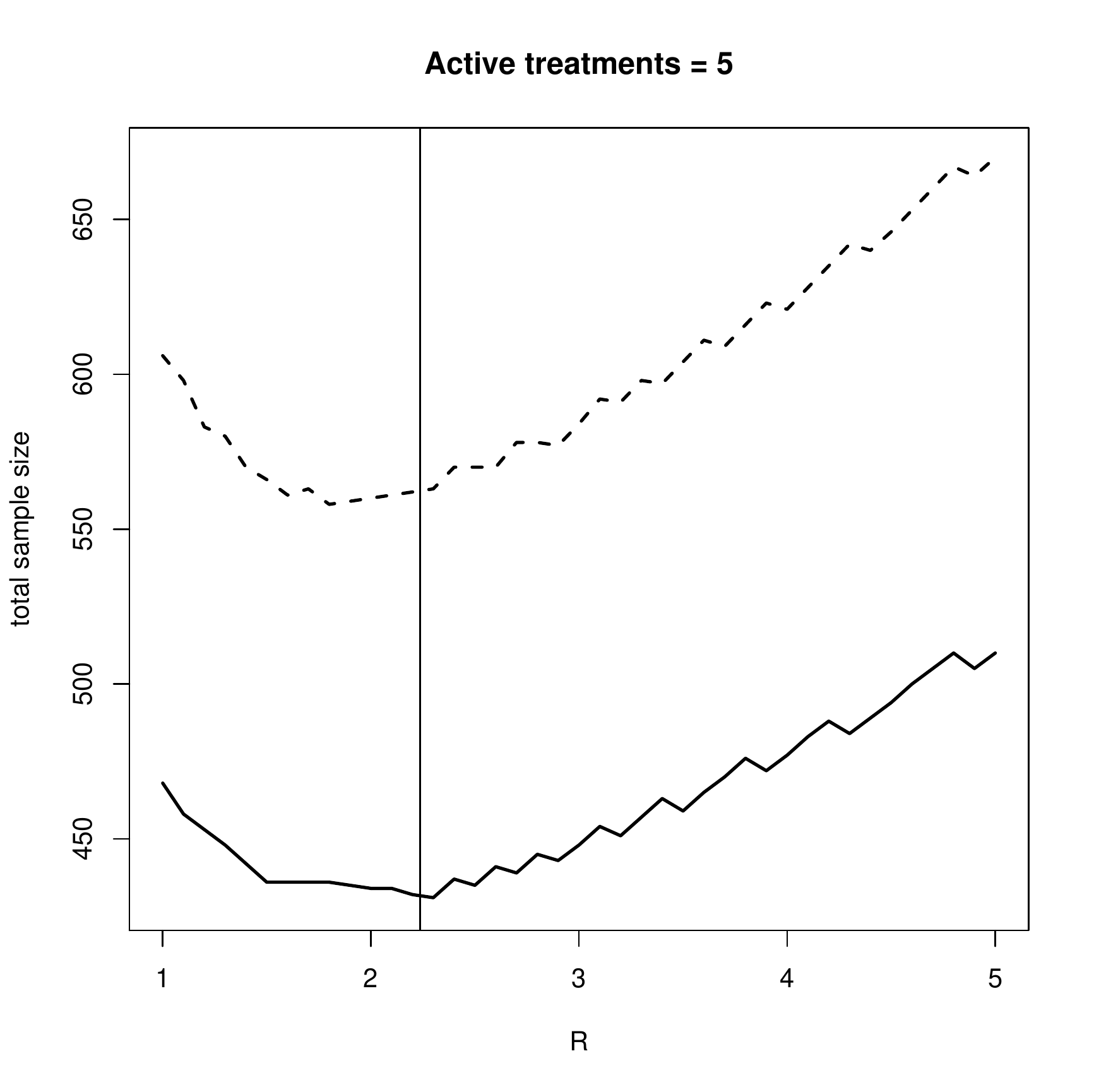}}
  \end{center}
  \caption{Change in total sample size for power $ = 0.8, 0.9$ and $K = 2, 3, 4, 5$, with fixed arguments $\alpha = 0.05, \sigma = 1, \delta = 0.5,$ and $\delta_0 = 0.125$}
  \label{fig:beta}
  \end{figure}  

  \begin{table}[htb]
	\centering
		\begin{tabular}{cc|c|c c c c}
		& & \textbf{Allocation ratio }& \multicolumn{4}{c}{\textbf{Active treatments } $\mathbf{K}$}\\
		 & & $\mathbf{R:1}$ & \multicolumn{1}{c}{2 } & \multicolumn{1}{c}{3 }& \multicolumn{1}{c}{4 }& \multicolumn{1}{c}{5 } \\
	     \hline
	     \hline
	      &&&&&&\\
	  %  & & $1 - \beta$					 					& 0.9		& 0.9		& 0.9		& 0.9\\
	    % & & $\mathbf{R:1}$ 												&			&			&			&\\		
\multirow{15}{*}{$\mathbf{\alpha}$}& \multirow{3}{*}{\textbf{0.2}} & $\mathbf{N(1:1)}$	& 159 		& 244		& 335		& 432\\
	     &&\textbf{2:1}														& 1.08 (-13)	& 1.05 (-11)	&1.00 (-1)	&0.99 (5)\\
	     &&$\mathbf{R_{OP}:1}$												& 0.99 (2)	& 0.99 (3)	&0.98 (6)		&0.96 (16)\\
	     &&&&&&\\
	      					& \multirow{3}{*}{\textbf{0.1}} & $\mathbf{N(1:1)}$	& 204		& 304		& 410		& 516\\
	     &&$\mathbf{2:1}$													& 1.04 (-8)	&1.00 (-1)	&0.97 (14)	& 0.95 (26)\\
	     &&$\mathbf{R_{OP}:1}$												& 0.99 (2)	&0.98 (7)		&0.95 (19)	& 0.95 (28)\\
 &&&&&&\\
	      					& \multirow{3}{*}{\textbf{0.05}} & $\mathbf{N(1:1)}$	& 249		& 364		& 485		& 606\\
	     &&$\mathbf{2:1}$													& 1.03 (-7)	& 0.98 (9)	& 0.94 (29)	&0.92 (46)\\
	     &&$\mathbf{R_{OP}:1}$   												& 0.99 (2)	& 0.96 (14)	&0.94 (30)	&0.92 (48)\\
 &&&&&&\\
	      					& \multirow{3}{*}{\textbf{0.025}} & $\mathbf{N(1:1)}$	& 297		& 428		& 560		& 702\\
	     &&$\mathbf{2:1}$													&1.01 (-3)	&0.96 (18)	&0.93 (38)		&0.91 (65)\\
	     &&$\mathbf{R_{OP}:1}$												&0.97 (8)		&0.95(23)		&0.93 (41)	&0.90 (70)\\
		\end{tabular}
	\caption{Reduction in total sample size  achieved by choosing a $2:1$ or $R_{OP}:1$ allocation ratio, as a proportion of total sample size for a trial using a $1:1$ ratio (actual reduction in brackets). Other arguments used: Power $= 0.9, \sigma = 1, \delta = 0.5, \delta_0 = 0.125$.}
	\label{tab:gains22}
\end{table}

% OLD CAPTION: Reduction in sample size achieved by choosing $R=2$ or $R=R_{OP}$, compared to $R=1$, for various values of $\alpha$ and number of active treatments. Reduction given by proportion and number (where proportion $>1$ indicates \textit{increase} in sample size).

%Reduction in sample size achieved by choosing $R=2$ or $R=R_{OP}$, compared to $R=1$, for various values of $\alpha$ and number of active treatments. Reduction given by proportion and number (where proportion $>1$ indicates \textit{increase} in sample size).

The reduction in total sample size achieved by choosing the optimal allocation ratio over the standard $1:1$ ratio is shown in Table \ref{tab:gains22}, as are those for choosing $2:1$. The change in total sample size is reported as a proportion of total sample size for the $1:1$ case. This proportional reduction in total sample size increases as the number of active treatments increases, and also as the type I error requirement becomes more strict. For the arguments considered, total sample size is only reduced by more than $5\%$ when the number of active treatments is $\geq 4$ and the type I error probability is $\leq 0.05$. In such circumstances, similar reductions can be made by simply choosing a more conventional $2:1$ allocation ratio. Varying power, standard deviation, $\delta$ and $\delta_0$ showed no proportional decrease in total sample size; a comparison may be made with Table \ref{tab:2}, which shows no proportional change in total sample size with power.\\% (see Appendix D).\\

  \begin{table}[htb]
	\centering
		\begin{tabular}{cc|c|c c c c}
		& & \textbf{Allocation ratio }& \multicolumn{4}{c}{\textbf{Active treatments } $\mathbf{K}$}\\
		 & & $\mathbf{R:1}$ & \multicolumn{1}{c}{\textbf{2} } & \multicolumn{1}{c}{\textbf{3} }& \multicolumn{1}{c}{\textbf{4}}& \multicolumn{1}{c}{\textbf{5}} \\
	     \hline
	     \hline
	      &&&&&&\\
	  %  & & $1 - \beta$					 					& 0.9		& 0.9		& 0.9		& 0.9\\
	    % & & $\mathbf{R:1}$ 												&			&			&			&\\		
\multirow{5}{*}{$\mathbf{1-\boldsymbol\beta}$}& \multirow{2}{*}{\textbf{0.8}} & $\mathbf{N(1:1)}$	& 186 		& 276		& 370		& 468\\
	    % &&\textbf{2:1}													& 1.08 (-13)	& 1.05 (-11)	&1.00 (-1)	&0.99 (5)\\
	     &&$\mathbf{R_{OP}:1}$												& 0.99 (2)	& 0.96 (10)	&0.94 (22)	&0.92 (37)\\
	     &&&&&&\\
	      					& \multirow{2}{*}{\textbf{0.9}} & $\mathbf{N(1:1)}$	& 249		& 364		& 485		& 606\\
	     %&&\textbf{2:1}													& 1.04 (-8)	&1.00 (-1)	&0.97 (14)	& 0.95 (26)\\
	     &&$\mathbf{R_{OP}:1}$												& 0.99 (2)	&0.96 (14)	&0.94 (30)	& 0.92 (48)\\
		\end{tabular}
	\caption{Reduction in total sample size achieved by choosing an $R_{OP}:1$ allocation ratio, as a proportion of total sample size for a trial using $1:1$ ratio (actual reduction in brackets). Other arguments used:  $\alpha = 0.05, \sigma = 1, \delta = 0.5, \delta_0 = 0.125$.}
	\label{tab:2}
\end{table}
 
%In Figure (\ref{fig:a}), the optimal allocation ratio is often slightly less than $\sqrt{K}$, the optimal allocation ratio expected from (\ref{eq:xx}). This is characteristic of many sets of arguments (see Figures XXX in Appendix D), and can be attributed to the choice of fixed arguments and the fact that (\ref{eq:xx}) was the result of a simplification; it is not due to the fact that $ N $ was not calculated for exactly $R = \sqrt{K}$. Where it occurs, this difference between the optimal allocation ratio and $\sqrt{K}$ increases as the number of active treatments $K$ increases. [WHY?] What can also been seen is the ``jagged'' characteristic of the curves. This is due to the integer nature of sample sizes. 

%For all combinations of arguments investigated, the decrease in sample size achieved by choosing the exact optimal allocation ratio over $R=1$ was always $\leq 10\%$. However, in many instances this decrease equated to a ``saving'' of 30 or more patients.  In such instances, choosing $R=2$ may also result in notable decrease in sample size. Reductions achieved are greater - both in actual numbers and by proportion - as the number of active treatments increases and as the probability of type I error decreases (Table \ref{tab:1}). In contrast, varying power, standard deviation, $\delta$ and $\delta_0$ showed no proportional increase in gains - compare with Table \ref{tab:2}.\\% (see Appendix D).\\

\subsection{Other benefits}
%%%%%%%%%%%
%FOR DISCUSSION: Choosing a trial's allocation ratio in order to minimise sample size appears to give notable improvement only under certain conditions - namely, when the number of active treatments is not small (i.e. is $>3$) and the type I error requirement is strict ($\leq 0.05$). Even when these conditions are met, some may consider the reduction in sample size - $\leq 10\%$ in all cases considered here - insufficient to warrant the added complexity of a non-integer sample size. In such cases, an allocation ratio of $2:1$ will provide valuable benefit without a considerable increase in complexity.
%%%%%%%%%%%

It is possible then, to use allocation ratio to decrease total sample size, though this decrease perhaps only becomes worthwhile under certain conditions - namely, when the number of active treatments is not small (i.e., $K \geq 4$) and the type I error requirement is strict ($ \alpha \leq 0.05$). However, it may not always be necessary to decrease total sample size in order to effect an improvement in a trial. So far, it has been assumed that the active and control treatments are approximately equal in toxicity and cost, and so a decrease in total sample size was sought. However, the case may arise that at least one active treatment is considerably more toxic or expensive than the control treatment. Under these circumstances, a decrease in the number of patients on the active treatments may be beneficial. Noting again Figures \ref{fig:a} and \ref{fig:beta}, it can be seen that it may be possible to increase allocation ratio without greatly increasing total sample size. As the function \textit{find.all} is designed to find the minimum total sample size which maintains type I error and power (and as $N=Rn+Kn$), using \textit{find.all} to increase the allocation ratio $R$ without increasing the total sample size will result in a decrease in the number of patients on each active treatment $n$, without compromising type I error or power.\\

Suppose that a slight increase in total sample size $N$ - say, $3\%$ - would be deemed acceptable provided there is a considerable decrease in the number of patients on the active treatments. Define $R_{MAX}:= \text{max}\{R \} : (Rn + Kn)/(1n + Kn) \leq 1.03$ - that is, $R_{MAX}$ is the greatest allocation ratio which does not increase total sample size by more than $3\%$ (compared to the $1:1$ case).\\  

Table \ref{tab:maxR} shows the extent to which the allocation ratio may be increased without a considerable increase in total sample size. Table \ref{tab:maxR2} quantifies this by showing the reduction in sample size \textit{for each active treatment}, both by proportion and in patient numbers. As was the case for $R_{OP}$, the benefit of choosing $R_{MAX}:1$ over $1:1$ increases both as the number of active treatments increases, and as the type I error is made more strict. However, whereas choosing $R_{OP}:1$ resulted in considerable reductions only under very specific conditions, an allocation ratio of $R_{MAX}:1$ appears to lead to a reduction in sample size of over $20\%$ for each active treatment provided $\alpha \leq 0.1$ - a sensible maximum value if one hopes to avoid incorrectly proceeding to a large-scale phase III trial. Further, as the decrease in sample size is for each active treatment, the benefit is increased when it is desired to minimise the sample size of a number of active treatments. Similar, though slightly smaller, reductions in sample size result from choosing $R=2$.

%choosing $R_{MAX}$ over $R=1$ leads to interesting decreases in the number of patients receiving each active treatment.

%This would be achieved by increasing the

\begin{table}[htb]
	\centering
		\begin{tabular}{cc|cccc}
		& & \multicolumn{4}{c}{ \textbf{Active treatments} $\mathbf{K}$}\\
	     & & \multicolumn{1}{c}{\textbf{2} } & \multicolumn{1}{c}{\textbf{3} }& \multicolumn{1}{c}{\textbf{4}}& \multicolumn{1}{c}{\textbf{5}} \\
	     \hline
%	     & $\mathbf{1 - \beta}$ 			 			& 0.9	& 0.9	& 0.9	& 0.9\\
	     \hline
	     &&&&&\\
	     \multirow{7}{*}{$\mathbf{\boldsymbol\alpha}$} & \textbf{0.2}	& 1.7	& 1.9	& 2.2	& 2.7\\
	      &&&&&\\
	      & \textbf{0.1}							& 1.9	& 2.4	& 2.9	& 3.4\\
	      &&&&&\\
	      & \textbf{0.05}							& 2.0	& 2.6	& 3.3	& 4.0\\
	      &&&&&\\
	      & \textbf{0.025}							& 2.2	& 2.9	& 3.7	& 4.5\\
		\end{tabular}
	\caption{The maximum allocation ratio, $R_{MAX}$, that does not result in an increase in total sample size $ > 3\% $ compared to a trial using $1:1$ ratio.} %$(\sigma = 1, \delta = 0.5, \delta_0 = 0.125)$ }
	\label{tab:maxR}
\end{table}

  \begin{table}[htb]
	\centering
		\begin{tabular}{cc|c|c c c c}
		& & \textbf{Allocation ratio }& \multicolumn{4}{c}{\textbf{Active treatments } $\mathbf{K}$}\\
		 & & $\mathbf{R:1}$ & \multicolumn{1}{c}{2 } & \multicolumn{1}{c}{3 }& \multicolumn{1}{c}{4 }& \multicolumn{1}{c}{5 } \\
	     \hline
	     \hline
	      &&&&&&\\
	  %  & & $1 - \beta$					 					& 0.9		& 0.9		& 0.9		& 0.9\\
	    % & & $\mathbf{R:1}$ 												&			&			&			&\\		
\multirow{15}{*}{$\mathbf{\alpha}$}& \multirow{3}{*}{\textbf{0.2}} & $\mathbf{n(1:1)}$	& 53 		& 61			& 67 		& 72\\
	     &&$\mathbf{2:1}$														& *			& *			&0.84 (11)	&0.85 (11)\\
	     &&$\mathbf{R_{MAX}:1}$												& 0.83 (9)	& 0.84 (10)	&0.82 (12)	&0.81 (14)\\
	     &&&&&&\\
	      					& \multirow{3}{*}{\textbf{0.1}} & $\mathbf{n(1:1)}$	& 68			& 76 		& 82			& 86\\
	     &&$\mathbf{2:1}$													& *			&0.80 (15)	&0.80 (16)	& 0.81 (16)\\
	     &&$\mathbf{R_{MAX}:1}$												& 0.79 (14)	&0.76 (18)	&0.74 (21)	& 0.73 (23)\\
 &&&&&&\\
	      					& \multirow{3}{*}{\textbf{0.05}} & $\mathbf{n(1:1)}$	& 83			& 91 		& 97 		& 101\\
	     &&$\mathbf{2:1}$													& 0.77 (19)	& 0.78 (20)	& 0.78 (21)	&0.79 (21)\\
	     &&$\mathbf{R_{MAX}:1}$   											& 0.77 (19)	& 0.74 (24)	&0.70 (29)	&0.68 (32)\\
 &&&&&&\\
	      					& \multirow{3}{*}{\textbf{0.025}} & $\mathbf{n(1:1)}$	& 99 		& 107		& 112		& 117\\
	     &&$\mathbf{2:1}$													&0.76 (24)	&0.77 (25)	&0.78 (25)	&0.78 (26)\\
	     &&$\mathbf{R_{MAX}:1}$												&0.74 (26)	&0.69 (33)	&0.67 (37)	&0.65 (41)\\
		\end{tabular}
	\caption{Reduction in sample size achieved by choosing a $2:1$ or $R_{MAX}:1$ allocation ratio, as a proportion of sample size for a trial using $1:1$ ratio (actual reduction in brackets). (*) indicates that total sample size $N$ was increased by $>3\%$.}
	\label{tab:maxR2}
\end{table}

\subsection*{Example: Simplified STAMPEDE trial}

The MRC's STAMPEDE trial is a five-stage, six-arm clinical trial investigating the efficacy of three drugs across five active treatment arms for the treatment of prostate cancer \cite{stampede}. Multi-stage trials and survival outcomes have not been considered in this study, so we consider a highly simplified version of the STAMPEDE trial as an example: Consider a single-stage clinical trial with one control arm and five active treatment arms. The active treatments for three arms are new drugs $X, Y$ and $Z$ respectively, while the active treatments for the remaining two arms are combination treatments, $X+Y$ and $X+Z$. The required type I error is $\alpha = 0.013$ and power is $1-\beta = 0.85$, as for the STAMPEDE trial. For simplicity, $\delta$ and $\delta_0$ remain $0.5$ and $0.125$ respectively, while $\sigma$ is increased to $1.5$ to more closely reflect the large number of patients (3,100) estimated to be required for the STAMPEDE trial. Suppose that drugs $Y$ and $Z$ are particularly toxic or expensive, and so it is of interest to minimise the number of patients receiving these drugs without loss of type I error, power, or greatly increasing total sample size. The effect of choosing allocation ratios $1:1,\; 2:1$ and $R_{MAX}:1$ are shown in Table \ref{tab:mrc}. Note that the actual STAMPEDE trial uses a $2:1$ allocation ratio.\\

\begin{figure}[htb]
	\centering
		\includegraphics[width=0.5\textwidth]{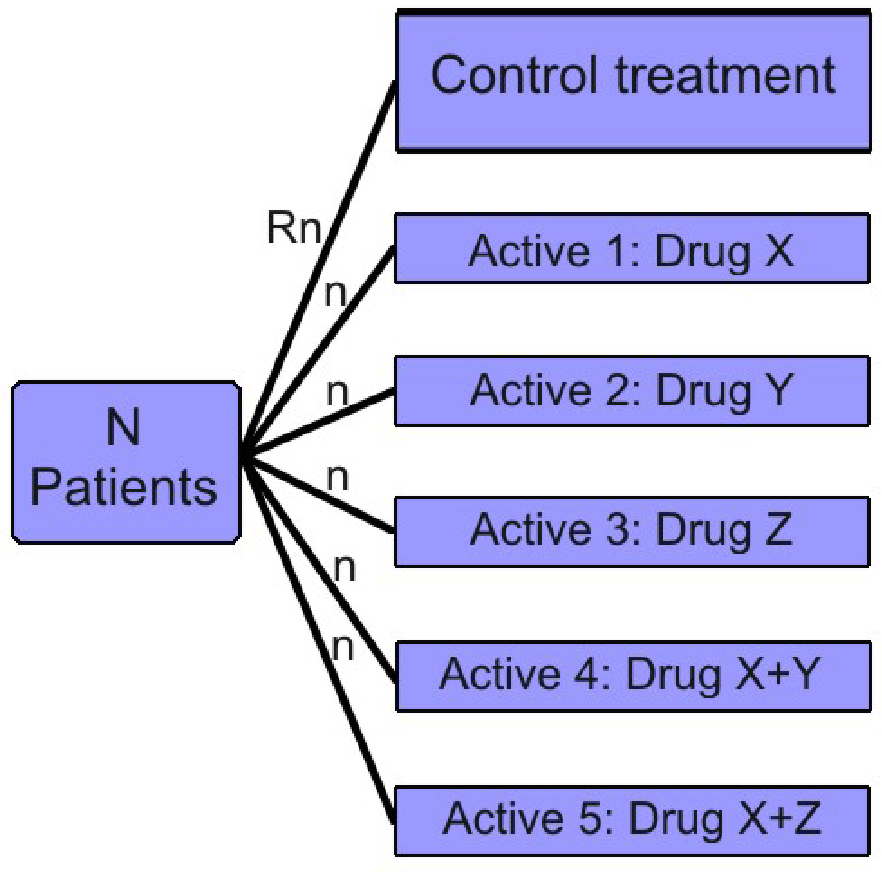}
	\caption{Example: STAMPEDE trial (simplified).}
	\label{fig:Stampede2}
\end{figure}

\begin{table}[htb]
	\centering
		\begin{tabular}{c|cccc}
		\textbf{Allocation ratio }& & & \textbf{Reduction per} & \textbf{Total reduction in patients }\\
		$\mathbf{R:1}$ & $\mathbf{N}$ & $\mathbf{n}$  & \textbf{ active treatment} & \textbf{ receiving drugs $Y$ or $Z$}\\
		\hline
		\hline
		&&&&\\
		$\mathbf{1:1}$ 			& 1560 & 260 & -- 		& -- \\
		$\mathbf{2:1}$ 			& 1393 & 199 & 61 (0.77) & $4\times61=244 (0.77)$\\
		$\mathbf{R_{MAX} (4.9):1}$ 	& 1614 & 163 & 97 (0.63) & $4\times97=388 (0.63)$\\
	\end{tabular}
	\caption{Example: Sample size reduction for each active treatment and combined reduction for treatment arms involving drugs $Y$ or $Z$, compared to a trial using $1:1$ ratio (proportion in brackets).}
	\label{tab:mrc}
\end{table}

In the example, as in Table \ref{tab:maxR2}, there is a considerable reduction in the sample size $n$ for each active treatment, for both $2:1$ and $R_{MAX}:1$ -- 97 per arm for the $R_{MAX}$ case, a reduction of almost $40\%$. As four of the five active treatment arms involve the toxic (or expensive) drugs $Y$ and $Z$, the total reduction in the number of patients being administered drugs $Y$ or $Z$ is $4 \times 97 = 388$. 

\section{Two-stage design}

\begin{figure}[ht]
  \begin{center}
   \subfigure[Active treatments = 2]{\label{fig:ak2}\includegraphics[scale=0.41]{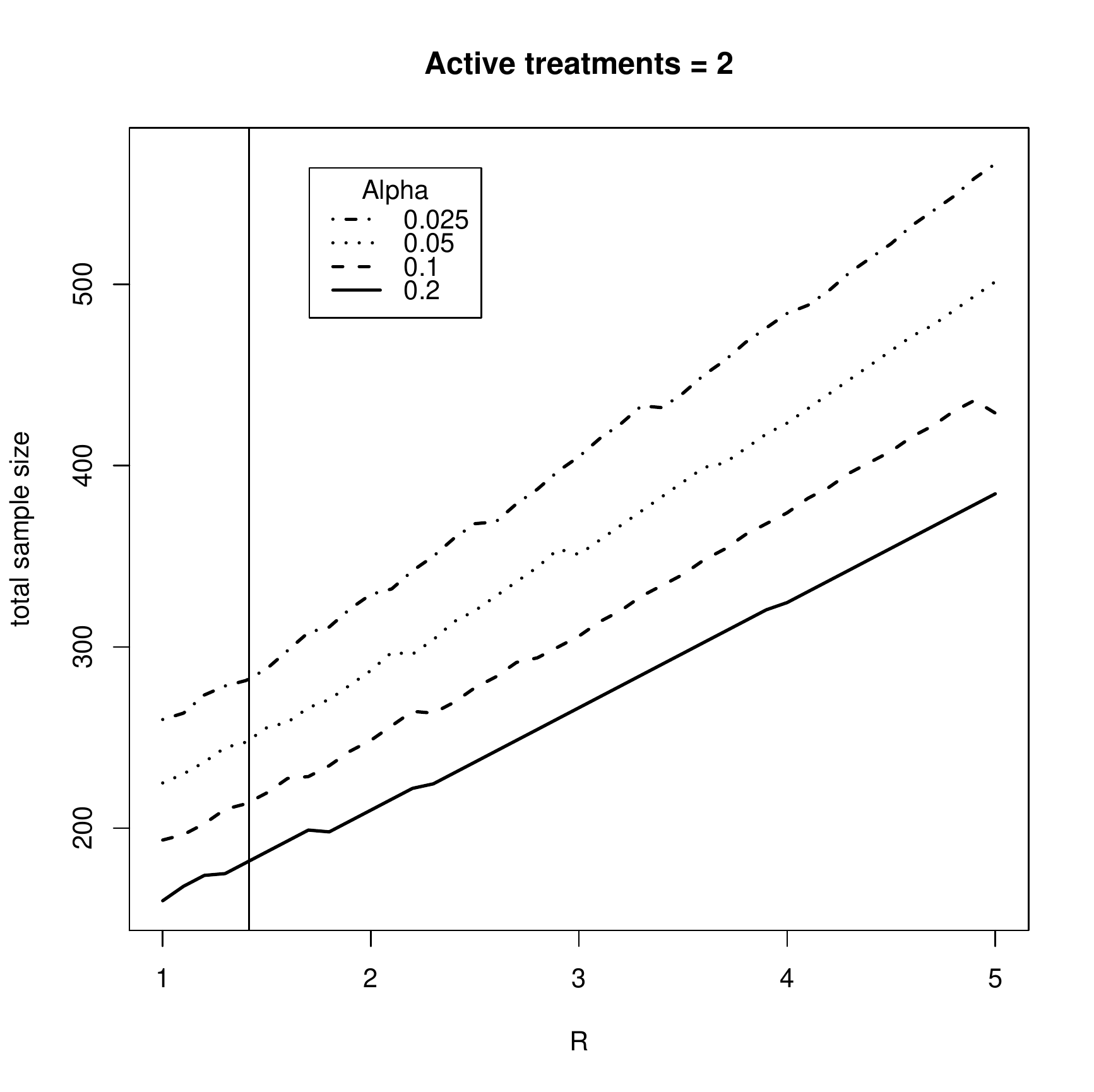}}
    \subfigure[Active treatments = 3]{\label{fig:ak3}\includegraphics[scale=0.41]{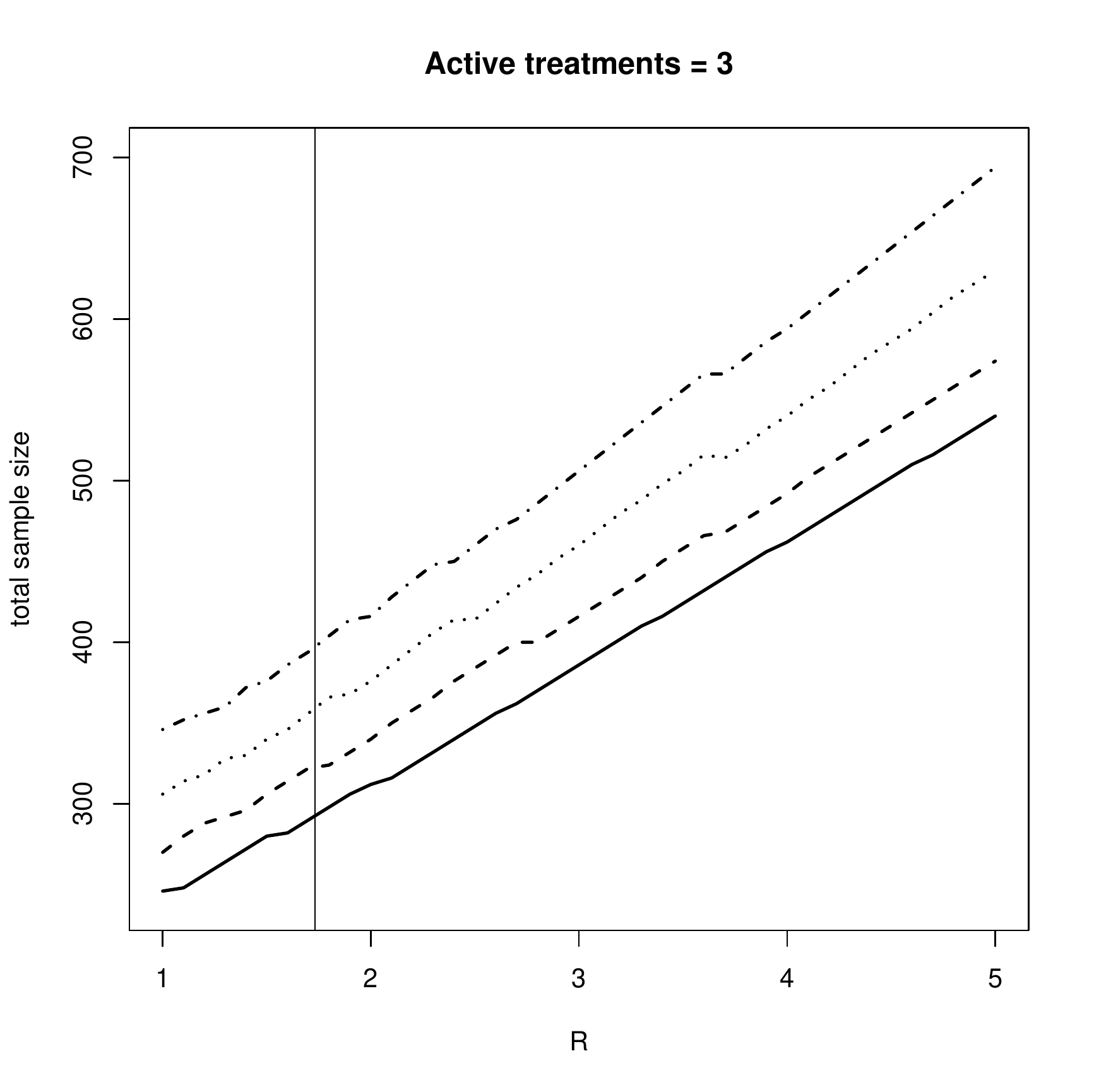}}
     \subfigure[Active treatments = 4]{\label{fig:ak4}\includegraphics[scale=0.41]{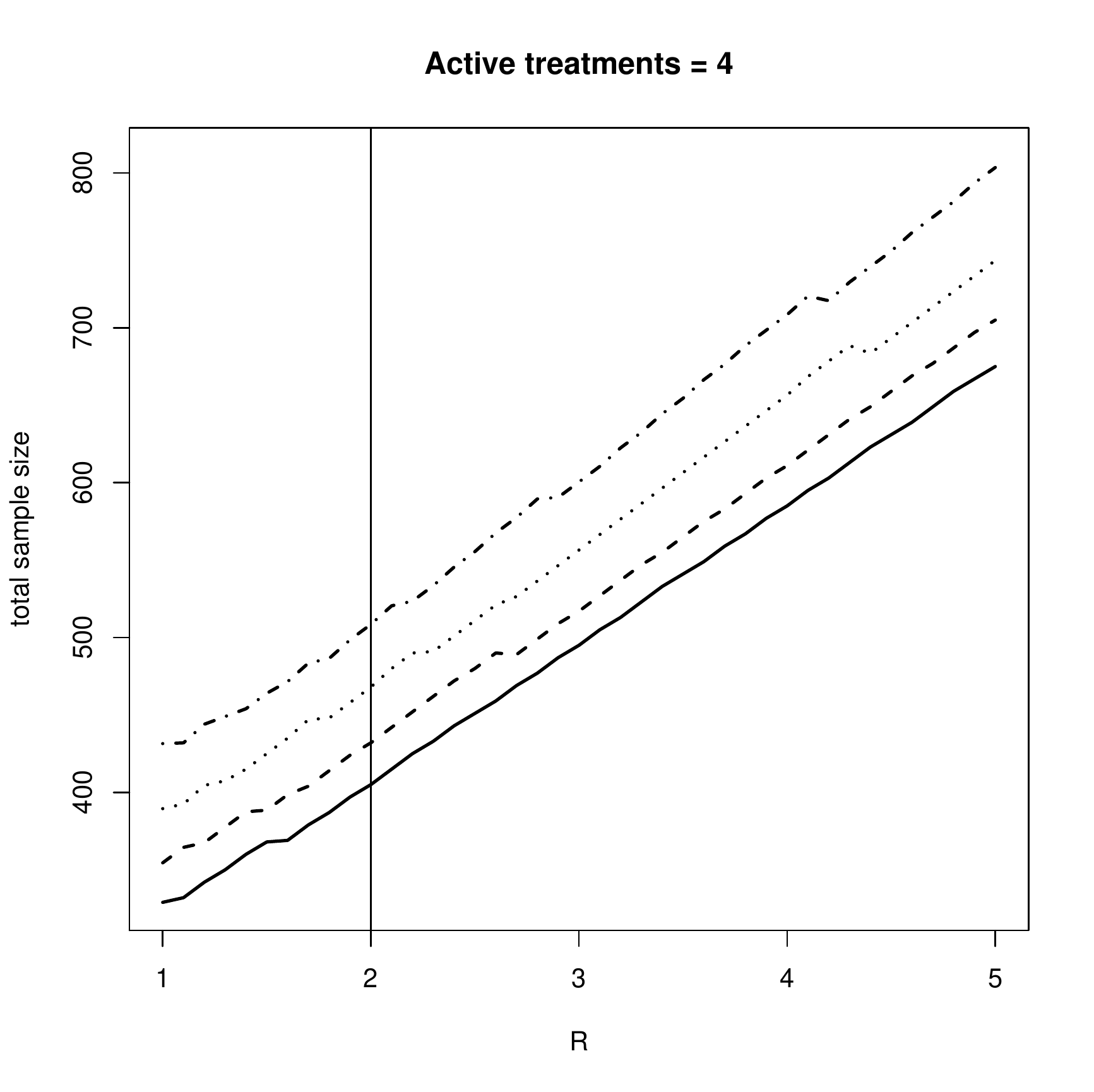}}
      \subfigure[Active treatments = 5]{\label{fig:ak5}\includegraphics[scale=0.41]{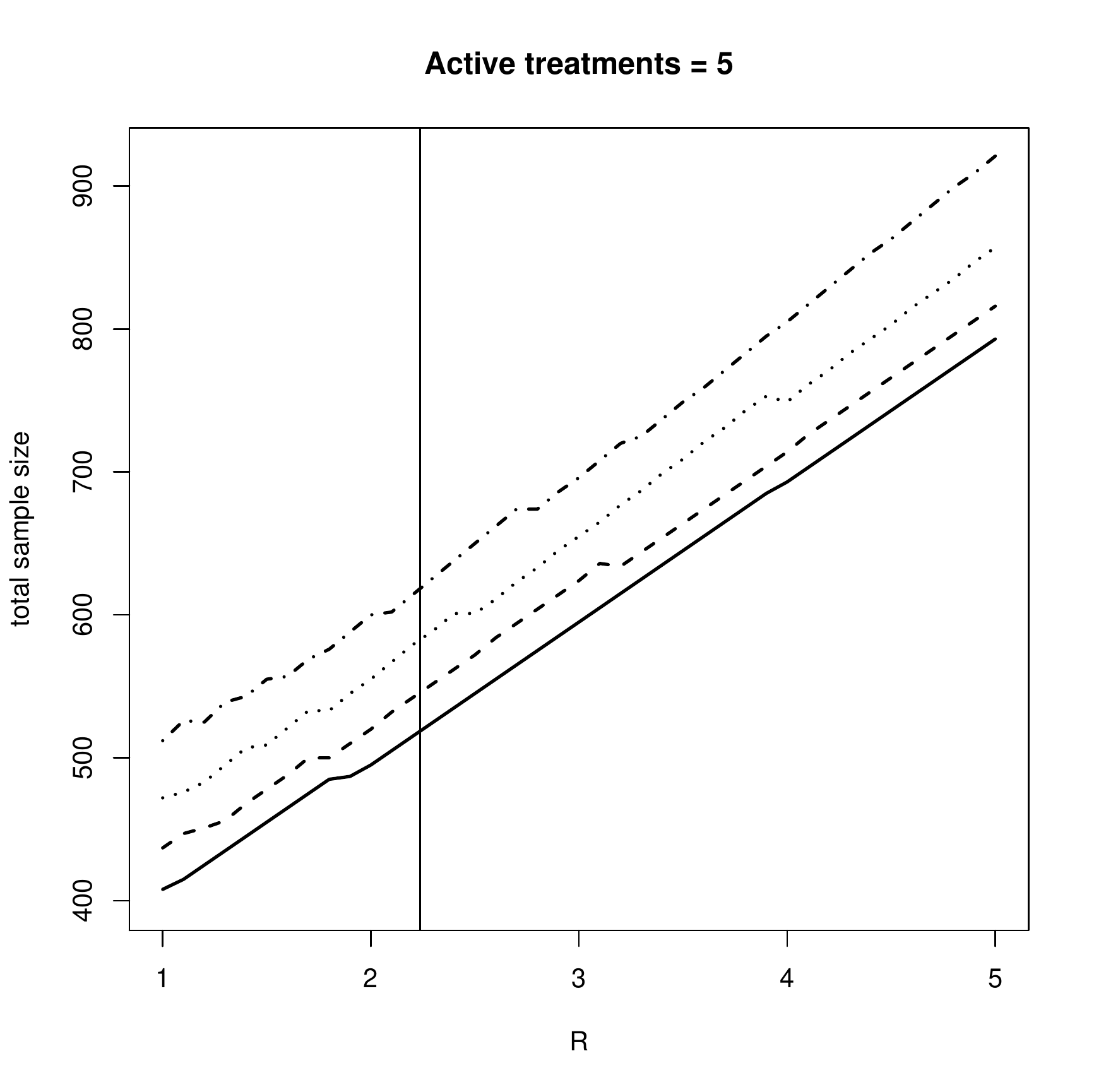}}
  \end{center}
  \caption{Two-stage design: Change in total sample size for $\alpha = 0.025, 0.05, 0.1, 0.2$ and $K = 2, 3, 4, 5$, with fixed arguments power $= 0.9, \sigma = 1, \delta = 0.5,$ and $\delta_0 = 0.125$}
  \label{fig:aselect}
  \end{figure}

\begin{figure}[ht]
  \begin{center}
   \subfigure[Active treatments = 2]{\label{fig:betak2}\includegraphics[scale=0.41]{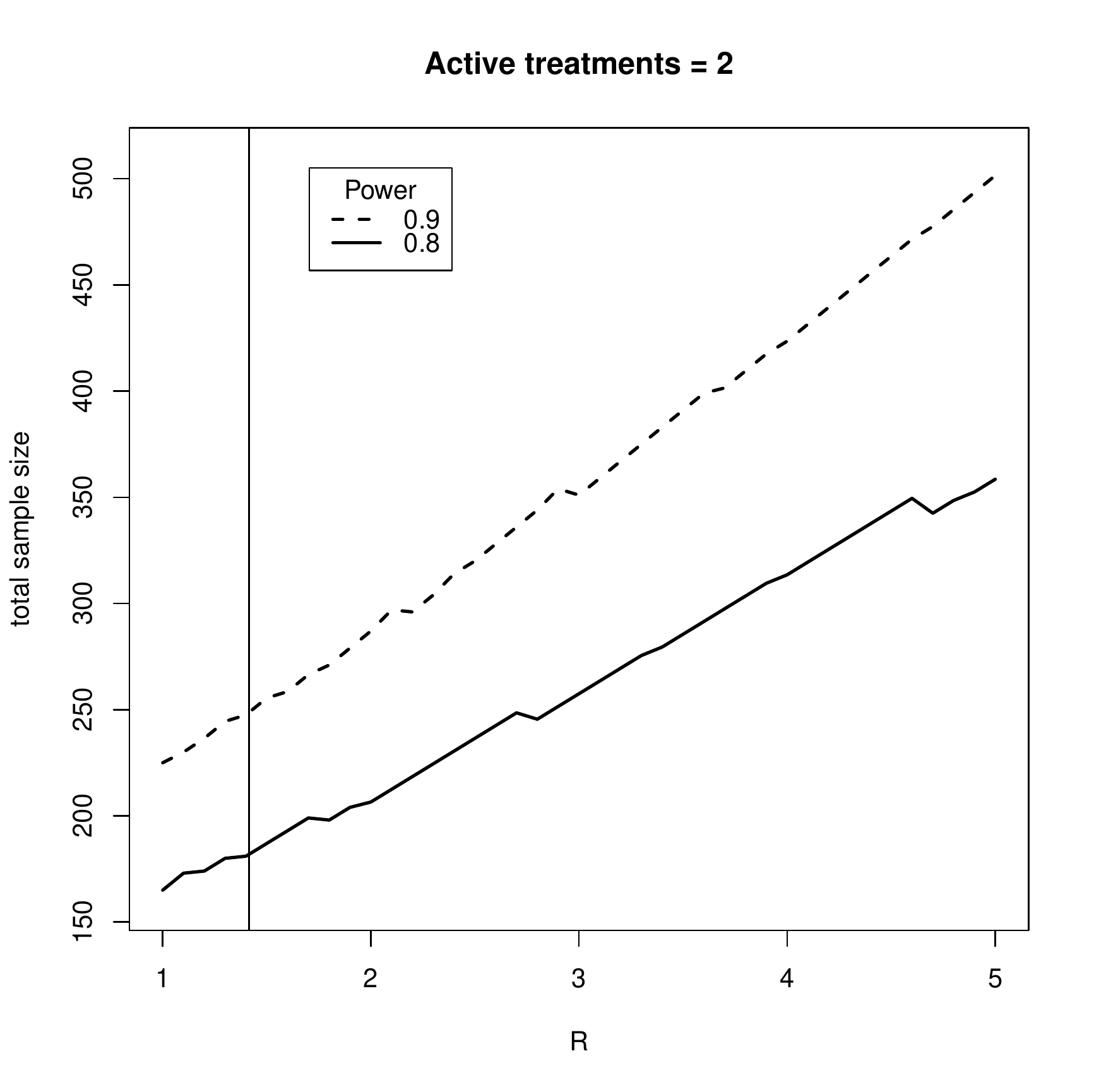}}
    \subfigure[Active treatments = 3]{\label{fig:betak3}\includegraphics[scale=0.41]{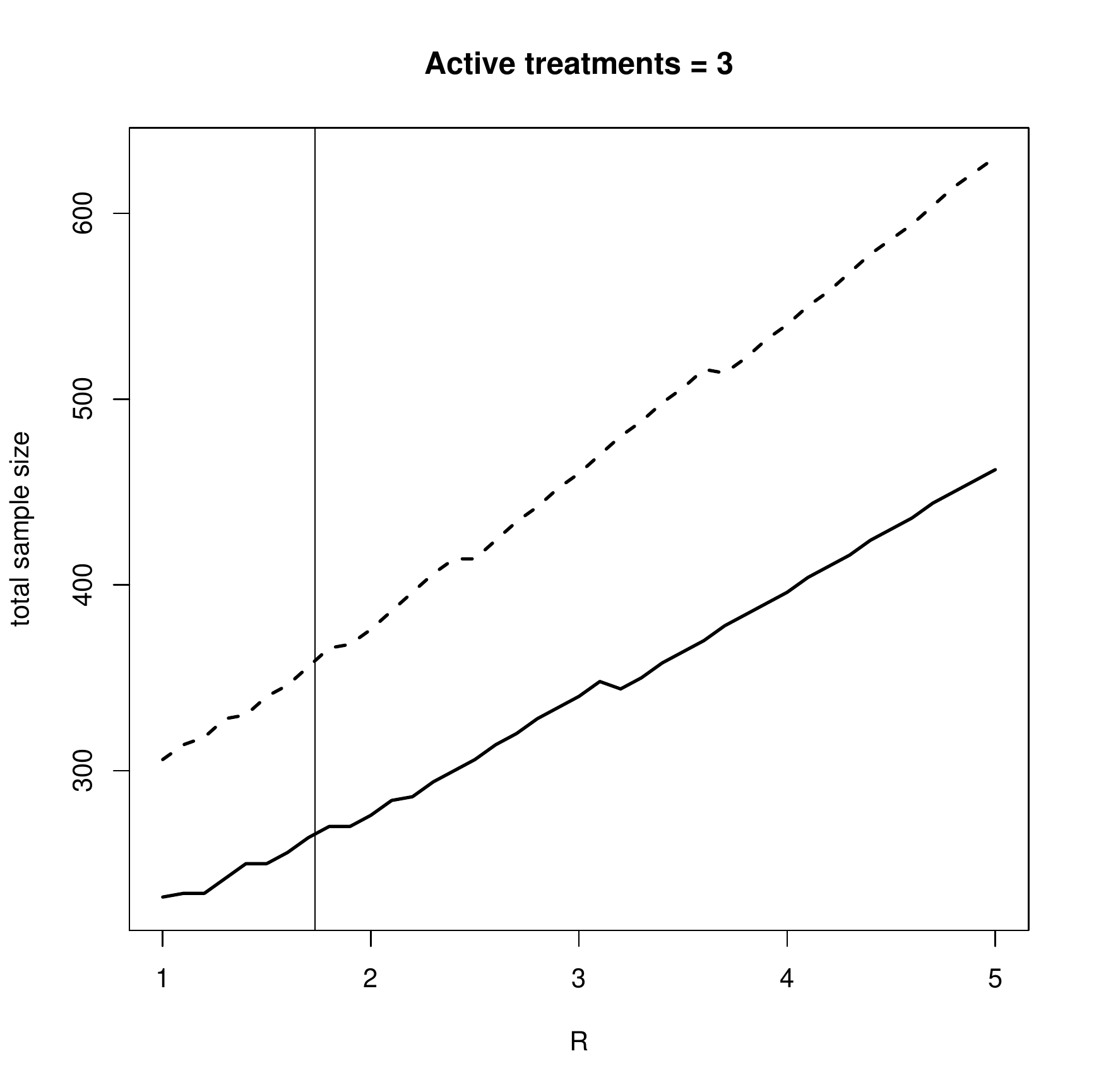}}
     \subfigure[Active treatments = 4]{\label{fig:betak4}\includegraphics[scale=0.41]{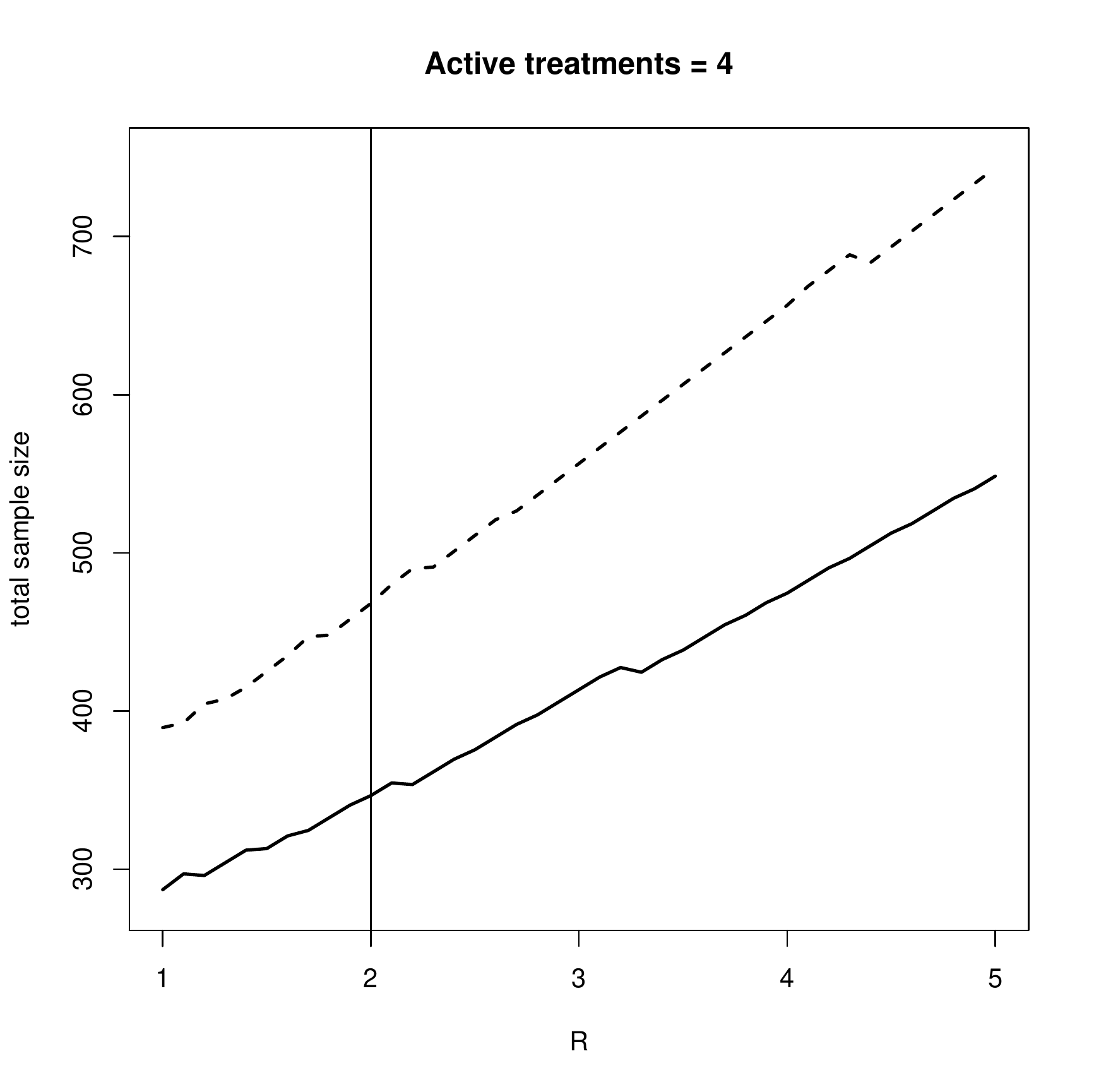}}
      \subfigure[Active treatments = 5]{\label{fig:betak5}\includegraphics[scale=0.41]{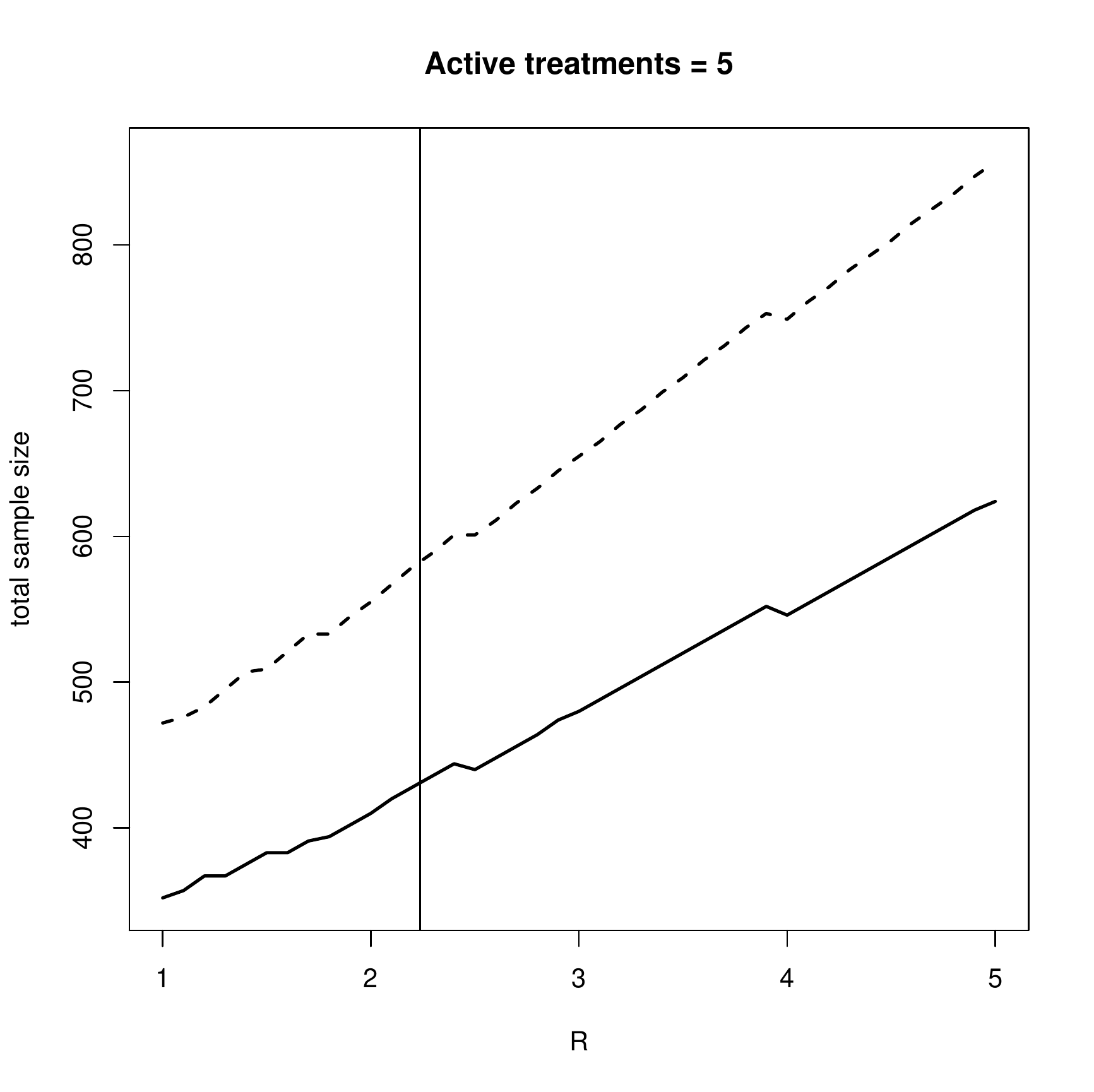}}
  \end{center}
  \caption{Two-stage design: Change in total sample size for power $ = 0.8, 0.9$ and $K = 2, 3, 4, 5$, with fixed arguments $\alpha = 0.05, \sigma = 1, \delta = 0.5,$ and $\delta_0 = 0.125$}
  \label{fig:betaselect}
  \end{figure}

Figures \ref{fig:aselect} and \ref{fig:betaselect} show how total sample size changes with allocation ratio for the two-stage, ``select only'' case. The relationship appears to be linear, with total sample size increasing with allocation ratio. Similarly to the one-stage design (Figures \ref{fig:a} and \ref{fig:beta}), the plots are jagged due to rounding by the function to ensure that the sample sizes remain integers. In almost all cases examined, the optimal allocation was 1. When this was not the case, the optimal allocation was 1.1, and resulted in a total sample size reduction of just $1\%$. Increasing the allocation ratio can cause sharp increases in total sample size. For example, for default arguments $\alpha = 0.05,$ power $= 0.9, \sigma = 1, \delta = 0.5$ and $\delta_0 = 0.125$, using a $2:1$ ratio results in a $28\%$ increase in total sample size, for $K=2$. Thus, there is little to be gained by varying allocation ratio in a trial using such a design.\\

\clearpage 

\chapter{Discussion}

For both the one- and two-stage designs presented, I have shown how the smallest total sample size changes as allocation ratio is varied. In the one-stage design, it was expected that the optimal allocation ratio $R_{OP}$ - that which minimises tss - would be $\sqrt{K}:1$, as suggested by Dunnett \cite{dunnett}. It was also expected that choosing $R_{OP}:1$ would result in a considerable reduction in total sample size compared to using a $1:1$ ratio, with the benefit increasing as the number of active treatments $K$ increased (as $\sqrt{K}$ moves further from 1 as $K$ increases). In reality, the optimal allocation is not strictly $\sqrt{K}:1$, as this value was derived using a simplification of the design (see Methods), and the true design is more complex, making the optimal allocation ratio dependent on more than simply the number of active treatments. The benefit of choosing $R_{OP}$ did increase as the number of active treatments increased and as the type I error requirement was made more strict. Further, the reductions possible are generally small, only becoming $>5\%$  under specific conditions - to wit, $K \geq 4$ and $\alpha \leq 0.05$. Even when these conditions are met, it was found that reductions of a similar magnitude by simply using an AR of $2:1$. Choosing this allocation ratio also avoids any additional complexity which may arise from using non-integer AR s. Also, while it may be possible to find an explicit analytic solution for the optimal allocation ratio using equations (\ref{eq:11}) and (\ref{eq:t22}), it seems likely that finding such a solution would be convoluted and, given these findings, of little interest. Therefore, if one wishes to reduce total sample size for a one-stage multi-arm clinical trial, it seems wise to consider using a $2:1$ allocation ratio, and only in the case that the above conditions are met.\\

For the one-stage design, it was noted from Figures \ref{fig:a} and \ref{fig:beta} that allocation ratio could be increased, often doubled or more, without increasing in total sample size by more than $3\%$ compared to that for a $1:1$ ratio. Increasing allocation ratio without increasing tss causes the number of patients on each active treatment to decrease. As when considering reductions in tss , the benefit of choosing an increased AR is increased both as the number of active treatments is increased and as the type I error requirement is made more strict. Indeed, in the example used, where $K=5$ and $\alpha=0.013$, the decrease in sample size on each active treatment arm was almost $40\%$ compared to using a $1:1$ ratio. This may be of use when designing a clinical trial for which there are a number of active treatments which are particularly toxic, expensive, or for some other reason it is desired to decrease the number of patients receiving such treatments. Peto \cite{peto} and Pocock \cite{pocock} both advocated the use of unequal allocation ratios for reasons of ethics or finance, though they have done so with a view to decreasing the number of patients receiving the control treatment - that is, using an AR of $R:1$ where $R<1$. Given more time, this notion would have been explored, and the results made clear. Purely from observation of Figures \ref{fig:a} and \ref{fig:beta}, it seems that it would be possible to choose $R<1$ without major increases in tss compared to using a $1:1$ ratio - particularly when the number of active treatments is low and the type I error requirement is less strict.\\
 
%total sample size and allocation ratio had an approximately quadratic relationship 

%that the optimal allocation ratio may be calculated, and its improvement in terms of total sample size over using a $1:1$ ratio evaluated. 

%%%%%\input{appx}
%\input{appx2}
%\input{R2}


\begin{thebibliography}{9}

\bibitem{EFPIA}
  European Federation of Pharmaceutical Industries and Associations,
  \emph{The Pharmaceutical Industry in Figures - 2011 update}.
  \begin{verbatim}{http://www.efpia.org}
  \end{verbatim} Accessed August 2011
  	  
\bibitem{kola}
  Kola I, Landis J,
  \emph{Can the pharmaceutical industry reduce attrition rates?}.
  Nature Reviews Drug Discovery 3, 711-716,
  August 2004
  
  	  
\bibitem{parmar}
  Parmar \emph{et al}
  \emph{Speeding up the Evaluation of New Agents in Cancer}.
  J Natl Cancer Inst;100: 1204 - 1214
  2008
  	  
\bibitem{stampede}
  Sydes \emph{et al} 
  \emph{Issues in applying multi-arm multi-stage methodology to a clinical trial in prostate cancer: the MRC STAMPEDE trial}.
  Trials, 10:39 doi:10.1186/1745-6215-10-39
  2009
  
  \bibitem{focus}
  Medical Research Council Clinical Trials Unit (Chief investigator: Prof Tim Maughan) 
  \emph{FOCUS-3: A study to determine the feasibility of molecular selection of therapy using KRAS, BRAF and topo-1 in patients with metastatic or locally advanced colorectal cancer}.
  \begin{verbatim}{http://tinyurl.com/mrcfocus3}
  \end{verbatim}
  Accessed August 2011
  
  \bibitem{icon6}
  Medical Research Council Clinical Trials Unit (Trial Manager: Monique Tomiczek)
  \emph{ICON6: A double-blind, placebo-controlled, three arm randomised multi-centre GCIG trial of AZD2171 in patients with ovarian cancer}.
  \begin{verbatim}{http://www.icon6.org}
  \end{verbatim}
   Accessed September 2011
   
 \bibitem{dunnett}
   C. W. Dunnett
  \emph{New Tables for Multiple Comparisons with a Control}.
  Biometrics (Sep. 1964), Vol. 20, No. 3, pp. 482-491
  
  \bibitem{thall}
   Thall \emph{et al}
  \emph{Two-stage selection and testing designs for comparative clinical trials}.
  Biometrika (1988), 75, 2, pp. 303-10
  
  \bibitem{R}
   R Development Core Team (2010). 
   \emph{R: A language and environment for
  statistical computing.} R Foundation for Statistical Computing,
  Vienna, Austria. ISBN 3-900051-07-0, URL 
  \begin{verbatim}{http://www.R-project.org}
\end{verbatim}
    
    \bibitem{stall}
   Stallard, N. and Todd, S. (2003) 
  \emph{Sequential designs for phase III clinical trials
incorporating treatment selection.}.
   Statistics in Medicine 22, 689-703.
    
  \bibitem{johnjack}
   Whitehead, J. and Jaki T. (2009) 
  \emph{One- and two-stage design proposals for a phase II
trial comparing three active treatments with control using an ordered categorical endpoint}.
Statistics in Medicine 28, 828-847

	\bibitem{amy}
	Whitehead, A. (2010)
	\emph{Designing an early phase clinical trial for a compound to treat uveal melanoma}.
	MSc dissertation, Lancaster university  
  
  \bibitem{peto}
   R. Peto, M. C. Pike \emph{et al}
   \emph{Design and analysis of randomized clinical trials requiring prolonged observation of each patient}.
   Br. J. Cancer (1976) 34, 585
  
  \bibitem{pocock}
   Pocock S. J.
  \emph{Allocation of Patients to Treatment in Clinical Trials}.
  Biometrics, Vol. 35, No. 1, Perspectives in Biometry (Mar., 1979), pp. 183-197
  
  
   
\end{thebibliography}
\end{document}